\documentclass[longauth]{aa}  

\usepackage{graphicx}
\usepackage[version=4]{mhchem}
\usepackage{txfonts}
\usepackage{subfigure}
\usepackage{orcidlink}
\usepackage{academicons}
\usepackage{tablefootnote}
\usepackage{gensymb}
\usepackage[normalem]{ulem}
\makeatletter
\renewcommand*\aa@pageof{, page \thepage{} of \pageref*{LastPage}}

\begin{document}

   \title{MINDS. Abundant water and varying C/O across the disk of Sz~98 as seen by JWST/MIRI}

   \author{Danny Gasman
          \inst{1}*\orcidlink{0000-0002-1257-7742}
          \and
          Ewine F. van Dishoeck\inst{2,3}\orcidlink{0000-0001-7591-1907}
          \and
          Sierra L. Grant\inst{3}\orcidlink{0000-0002-4022-4899}
          \and 
          Milou Temmink\inst{2}\orcidlink{0000-0002-7935-7445}
          \and
          Beno\^{i}t Tabone\inst{4}\orcidlink{0000-0002-1103-3225}
          \and
          Thomas Henning\inst{5}\orcidlink{0000-0002-1493-300X}
          \and
          Inga Kamp\inst{6}\orcidlink{0000-0001-7455-5349}
          \and
          Manuel G\"udel\inst{5,7,8}\orcidlink{0000-0001-9818-0588}
          \and
          Pierre-Olivier Lagage\inst{9}
          \and
          Giulia Perotti\inst{5}\orcidlink{0000-0002-8545-6175}
          \and
          Valentin Christiaens\inst{10}\orcidlink{0000-0002-0101-8814}
          \and
          Matthias Samland\inst{5}\orcidlink{0000-0001-9992-4067}
          \and
          Aditya M. Arabhavi\inst{6}\orcidlink{0000-0001-8407-4020}
          \and
          Ioannis Argyriou\inst{1}\orcidlink{0000-0003-2820-1077}
          \and
          Alain Abergel\inst{4}
          \and
          Olivier Absil\inst{10}\orcidlink{0000-0002-4006-6237}
          \and
          David Barrado\inst{11}\orcidlink{0000-0002-5971-9242}
          \and
          Anthony Boccaletti\inst{12}
          \and
          Jeroen Bouwman\inst{5}\orcidlink{0000-0003-4757-2500}
          \and
          Alessio Caratti o Garatti\inst{13,14}\orcidlink{0000-0001-8876-6614}
          \and
          Vincent Geers\inst{15}\orcidlink{0000-0003-2692-8926}
          \and
          Adrian M. Glauser\inst{8}\orcidlink{0000-0001-9250-1547}
          \and
          Rodrigo Guadarrama\inst{7}
          \and
          Hyerin Jang\inst{16}
          \and
          Jayatee Kanwar\inst{6,17,18}
          \and
          Fred Lahuis\inst{19}
          \and
          Maria Morales-Calder\'on\inst{11}
          \and
          Michael Mueller\inst{6}\orcidlink{0000-0003-3217-5385}
          \and
          Cyrine Nehm\'e\inst{9}
          \and
          G\"oran Olofsson\inst{20}\orcidlink{0000-0003-3747-7120}
          \and
          Eric Pantin\inst{9}
          \and
          Nicole Pawellek\inst{7,21}\orcidlink{0000-0002-9385-9820}
          \and
          Tom P. Ray\inst{14}\orcidlink{0000-0002-2110-1068}
          \and
          Donna Rodgers-Lee\inst{14}\orcidlink{0000-0002-0100-1297}
          \and
          Silvia Scheithauer\inst{5}\orcidlink{0000-0003-4559-0721}
          \and
          J\"urgen Schreiber\inst{5}
          \and
          Kamber Schwarz\inst{5}\orcidlink{0000-0002-6429-9457}
          \and
          Bart Vandenbussche\inst{1}\orcidlink{0000-0002-1368-3109}
          \and
          Marissa Vlasblom\inst{2}\orcidlink{0000-0002-3135-2477}
          \and
          L. B. F. M. Waters\inst{16,19}\orcidlink{0000-0002-5462-9387}
          \and
          Gillian Wright\inst{15}
          \and
          Luis Colina\inst{22}
          \and
          Thomas R. Greve\inst{23}\orcidlink{0000-0002-2554-1837}
          \and
          G\"oran \"Ostlin\inst{20}\orcidlink{0000-0002-3005-1349}
          }

   \institute{Institute of Astronomy, KU Leuven, Celestijnenlaan 200D, 3001 Leuven,             Belgium
         \and
            Leiden Observatory, Leiden University, 2300 RA Leiden, the Netherlands
        \and
            Max-Planck Institut f\"{u}r Extraterrestrische Physik (MPE), Giessenbachstr. 1, 85748, Garching, Germany
        \and
            Universit\'e Paris-Saclay, CNRS, Institut d’Astrophysique Spatiale, 91405, Orsay, France
        \and
            Max-Planck-Institut f\"{u}r Astronomie (MPIA), K\"{o}nigstuhl 17, 69117 Heidelberg, Germany
        \and
            Kapteyn Astronomical Institute, Rijksuniversiteit Groningen, Postbus 800, 9700AV Groningen, The Netherlands
        \and
            Dept. of Astrophysics, University of Vienna, T\"urkenschanzstr. 17, A-1180 Vienna, Austria
        \and
            ETH Z\"urich, Institute for Particle Physics and Astrophysics, Wolfgang-Pauli-Str. 27, 8093 Z\"urich, Switzerland
        \and
            Universit\'e Paris-Saclay, Universit\'e Paris Cit\'e, CEA, CNRS, AIM, F-91191 Gif-sur-Yvette, France
        \and
            STAR Institute, Universit\'e de Li\`ege, All\'ee du Six Ao\^ut 19c, 4000 Li\`ege, Belgium
        \and
            Centro de Astrobiolog\'ia (CAB), CSIC-INTA, ESAC Campus, Camino Bajo del Castillo s/n, 28692 Villanueva de la Ca\~nada,
            Madrid, Spain
        \and
            LESIA, Observatoire de Paris, Universit\'e PSL, CNRS, Sorbonne Universit\'e, Universit\'e de Paris, 5 place Jules Janssen, 92195 Meudon, France
        \and
            INAF – Osservatorio Astronomico di Capodimonte, Salita Moiariello 16, 80131 Napoli, Italy
        \and   
            Dublin Institute for Advanced Studies, 31 Fitzwilliam Place, D02 XF86 Dublin, Ireland
        \and
            UK Astronomy Technology Centre, Royal Observatory Edinburgh, Blackford Hill, Edinburgh EH9 3HJ, UK
        \and
            Department of Astrophysics/IMAPP, Radboud University, PO Box 9010, 6500 GL Nijmegen, The Netherlands
        \and
            Space Research Institute, Austrian Academy of Sciences, Schmiedlstr. 6, A-8042, Graz, Austria
        \and
            TU Graz, Fakultät für Mathematik, Physik und Geodäsie, Petersgasse 16 8010 Graz, Austria
        \and
            SRON Netherlands Institute for Space Research, PO Box 800, 9700 AV, Groningen, The Netherlands
        \and
            Department of Astronomy, Stockholm University, AlbaNova University Center, 10691 Stockholm, Sweden
        \and
            Konkoly Observatory, Research Centre for Astronomy and Earth Sciences, E\"otv\"os Lor\'and Research Network (ELKH), Konkoly-Thege Mikl\'os \'ut 15-17, H-1121 Budapest, Hungary
        \and
            Centro de Astrobiolog\'ia (CAB, CSIC-INTA), Carretera de Ajalvir, E-28850 Torrej\'on de Ardoz, Madrid, Spain
        \and
            DTU Space, Technical University of Denmark, Building 328, Elektrovej, 2800 Kgs. Lyngby, Denmark
        \\
            *\email{danny.gasman@kuleuven.be}
             }


   \date{Received 25 May 2023 / Accepted 23 September 2023}

  \abstract
   {The Mid-InfraRed Instrument (MIRI) Medium Resolution Spectrometer (MRS) on board the \textit{James Webb} Space Telescope (\textit{JWST}) allows us to probe the inner regions of protoplanetary disks, where the elevated temperatures result in an active chemistry and where the gas composition may dictate the composition of planets forming in this region. The disk around the classical T~Tauri star Sz~98, which has an unusually large dust disk in the millimetre with a compact core, was observed with the MRS, and we examine its spectrum here.}
   {We aim to explain the observations and put the disk of Sz~98 in context with other disks, with a focus on the \ce{H2O} emission through both its ro-vibrational and pure rotational emission. Furthermore, we compare our chemical findings with those obtained for the outer disk from Atacama Large Millimeter/submillimeter Array (ALMA) observations.}
   {In order to model the molecular features in the spectrum, the continuum was subtracted and local thermodynamic equilibrium (LTE) slab models were fitted. The spectrum was divided into different wavelength regions corresponding to \ce{H2O} lines of different excitation conditions, and the slab model fits were performed individually per region.}
   {We confidently detect \ce{CO}, \ce{H2O}, \ce{OH}, \ce{CO2}, and \ce{HCN} in the emitting layers. Despite the plethora of \ce{H2O} lines, the isotopologue H$_\text{2}^{\text{18}}$O is not detected. Additionally, no other organics, including \ce{C2H2}, are detected. This indicates that the C/O ratio could be substantially below unity, in contrast with the outer disk. The \ce{H2O} emission traces a large radial disk surface region, as evidenced by the gradually changing excitation temperatures and emitting radii. Additionally, the \ce{OH} and \ce{CO2} emission is relatively weak. It is likely that \ce{H2O} is not significantly photodissociated, either due to self-shielding against the stellar irradiation, or UV shielding from small dust particles. While \ce{H2O} is prominent and \ce{OH} is relatively weak, the line fluxes in the inner disk of Sz~98 are not outliers compared to other disks.}
   {The relative emitting strength of the different identified molecular features points towards UV shielding of \ce{H2O} in the inner disk of Sz~98, with a thin layer of \ce{OH} on top. The majority of the organic molecules are either hidden below the dust continuum, or not present. In general, the inferred composition points to a sub-solar C/O ratio ($<$0.5) in the inner disk, in contrast with the larger than unity C/O ratio in the gas in the outer disk found with ALMA.}


    \titlerunning{MINDS. Abundant water and varying C/O across the disk of Sz~98 as seen by JWST/MIRI}
    \authorrunning{D. Gasman et al.}
   \maketitle

%

\section{Introduction}
\label{sec:introduction}
The inner 0.1 to 10~au regions of protoplanetary disks are likely to be the cradle for terrestrial planets around low-mass stars. The high temperatures ($\ge 100$~K) and densities ($\ge 10^{8}$~cm$^{-3}$) in these regions, along with the locations of the H$_2$O and CO$_2$ snow lines and the presence of substructures (see e.g. \citealt{ref:23GrDiTa,ref:23TaBeDi,ref:23DiGrTa}; and \citealt{ref:14PoSaBe} for a review), dictate the composition of the gas and therefore the elemental abundances available to atmosphere formation of accreting planets.

The mid-infrared wavelength range observed by the Medium Resolution Spectroscopy \citep[MRS;][]{ref:15WePeGl,ref:23ArGlLa} mode of the Mid-InfraRed Instrument \citep[MIRI;][]{ref:15WrWrGo,ref:15RiWrBo,ref:23WrRiGl} on board the \textit{James Webb} Space Telescope \citep[\textit{JWST;}][]{ref:23RiPeMc} allows us to examine these inner regions of protoplanetary disks. Its wide wavelength range (4.9 to 28.1~$\mu$m) covers a variety of molecular features, including the large forest of \ce{H2O} lines: from the ro-vibrational bending mode from 5 to 8~$\mu$m, to the pure rotational lines around 10~$\mu$m and onwards \citep{ref:09MePoBl}. These lines are thought to probe radially different regions of the disk. Generally, the temperature of the gas probed decreases as the wavelength increases, likely corresponding to moving from the inner disk outwards \citep[e.g.][]{ref:17BaPoSa,ref:23BaPoPe}. The inner disk chemistry is now starting to be seen with MIRI/MRS \citep[e.g.][]{ref:23KoAbDi,ref:23GrDiTa,ref:23TaBeDi,ref:23KaHeAr,ref:23DiGrTa,ref:23PeChHe}.

Excitation of \ce{H2O} can occur due to collisions with H, H$_2$, He, and electrons; radiation from hot dust; photodesorption from dust grains; and chemical formation \citep{ref:09MePoBl,ref:13DiHeNe,ref:21DiKrMo}. Already in the era of \textit{Spitzer}, its InfraRed Spectrograph (IRS) unveiled a large diversity between the compositions of disks around T~Tauri stars \citep[e.g.][]{ref:10PoSaBl,ref:11CaNa,ref:14PoSaBe}. The samples examined by \citet{ref:11CaNa} and \citet{ref:13NaCaPo} showed that the ratio of \ce{HCN} to \ce{H2O} increases with increasing disk mass. Recently, \citet{ref:20BaPaBo} observed a similar correlation: the ratio of \ce{H2O} versus carbon-bearing molecules seems to correlate with disk size. This is interpreted as a smaller disk size indicating more efficient drift of icy pebbles, allowing the inner disk to be replenished with an ice reservoir that may sublimate. On the other hand, substructures in larger disks can prevent this transport. The disk of Sz~98, which is studied here, contains both of these features, being a large dust disk with rings, but also showing bright millimetre wavelength emission within several tens of au from the star, which we refer to as the central core \citep[e.g.][]{ref:18TeDiAn}.

A pathway for formation of \ce{H2O} in the gas phase is from its precursor OH: \ce{OH + H2 -> H2O + H}, which is most efficient at higher temperatures, typically $>250$~K \citep[e.g.][]{ref:09WoThKa,ref:09GlMeNa,ref:13DiHeNe}. In reverse, the \ce{OH} reservoir can be replenished again by photodissociation of \ce{H2O}, which puts \ce{OH} back into the cycle \citep[e.g.][]{ref:00HaHwYa,ref:00HaHe,ref:21TaHeDi}. The balance of the pebble drift and chemical formation from \ce{OH} producing \ce{H2O}, versus destruction by irradiation of the disk ultimately dictate the abundance of gaseous \ce{H2O} across the disk probed here in the infrared.

Sz~98 is a relatively cool, actively accreting classical T~Tauri star \citep{ref:08MeJoSp,ref:11MoOlVa} of spectral type K7 \citep{ref:17AlMaNa} with a disk amongst the 2\% largest and brightest dust disks in Lupus \citep{ref:18TeDiAn}. It has a luminosity $L_*=1.5$~$L_{\sun}$ and a mass $M_*=0.74$~$M_{\sun}$ \citep{ref:17AlMaNa} at a \textit{Gaia} distance of approximately 156~pc \citep{ref:gaia1,ref:gaia2}, and disk radii $R_{gas}$ and $R_{dust}$ of 360 and 180~au, respectively \citep{ref:18AnWiTr}. The mass of the disk is estimated to be 0.07~$M_{\sun}$ \citep{ref:19TeDiCa}. The star has an accretion rate and luminosity of $\dot{M}_{acc}\approx 10^{-7.6}$~$M_{\sun}\text{yr}^{-1}$ \citep{ref:17AlMaNa} and $L_{acc}\approx 10^{-0.72}$~$L_{\sun}$ \citep{ref:18NiAnAl}. In general the disk has not been found to have large substructures, aside from a small continuum break around 80~au and a ring around 90~au \citep{ref:17TaTeNa,ref:18TeDiAn,ref:19VaRuDi,ref:19MiFaVa}. Some evidence for additional ring-like structure has been found around $\sim$120~au \citep{ref:19TeDiCa}. Furthermore, based on scattered light images from the Spectro-Polarimetric High-contrast Exoplanet REsearch \citep[SPHERE;][]{ref:22GaDoGi} instrument, the disk's inner rim might be casting a uniform shadow on the outer disk. \citet{ref:19MiFaVa} suggest that the outer disk is depleted in volatile gaseous carbon and oxygen, and find signs for a gaseous C/O ratio larger than unity based on bright \ce{C2H} and faint \ce{^{13}CO} emission seen in Atacama Large Millimeter/submillimeter Array (ALMA) observations. The dust grains are expected to have grown at least up to millimetre sizes \citep{ref:07LoWrMa,ref:12UbMaWr}. The dust within 0.5~au from the star is likely warm (dust excess of $T\sim 600$~K) based on photometry with \textit{Spitzer} \citep{ref:10WaCiKo}. Sz~98 was previously observed with \textit{Spitzer}/IRS in 2008, but in its low resolution mode only. The higher sensitivity and spectral resolving resolution of the MRS ranging from a resolving power $\lambda / \Delta \lambda \approx 3500$ at shorter wavelengths to $\lambda / \Delta \lambda \approx 1500$ at the longer wavelengths \citep{ref:23JoAlSl} now allows us to detect \ce{H2O} in all of its bands.


Since we detect water across the full MIRI/MRS wavelength range, the focus of this paper is on the \ce{H2O}. The paper is structured as follows. The data reduction is described in Sect. \ref{sec:methods}, where we also discuss our modelling methods. The resulting spectrum and best model fits are shown in Sect. \ref{sec:results}. A discussion regarding the implications on the chemistry in the inner disk of Sz~98 and a comparison to other disks can be found in Sect. \ref{sec:discussion}. Finally, we summarise our conclusions in Sect. \ref{sec:conclusion}.

\begin{figure*}[t]
    \centering
    \includegraphics[width=\textwidth]{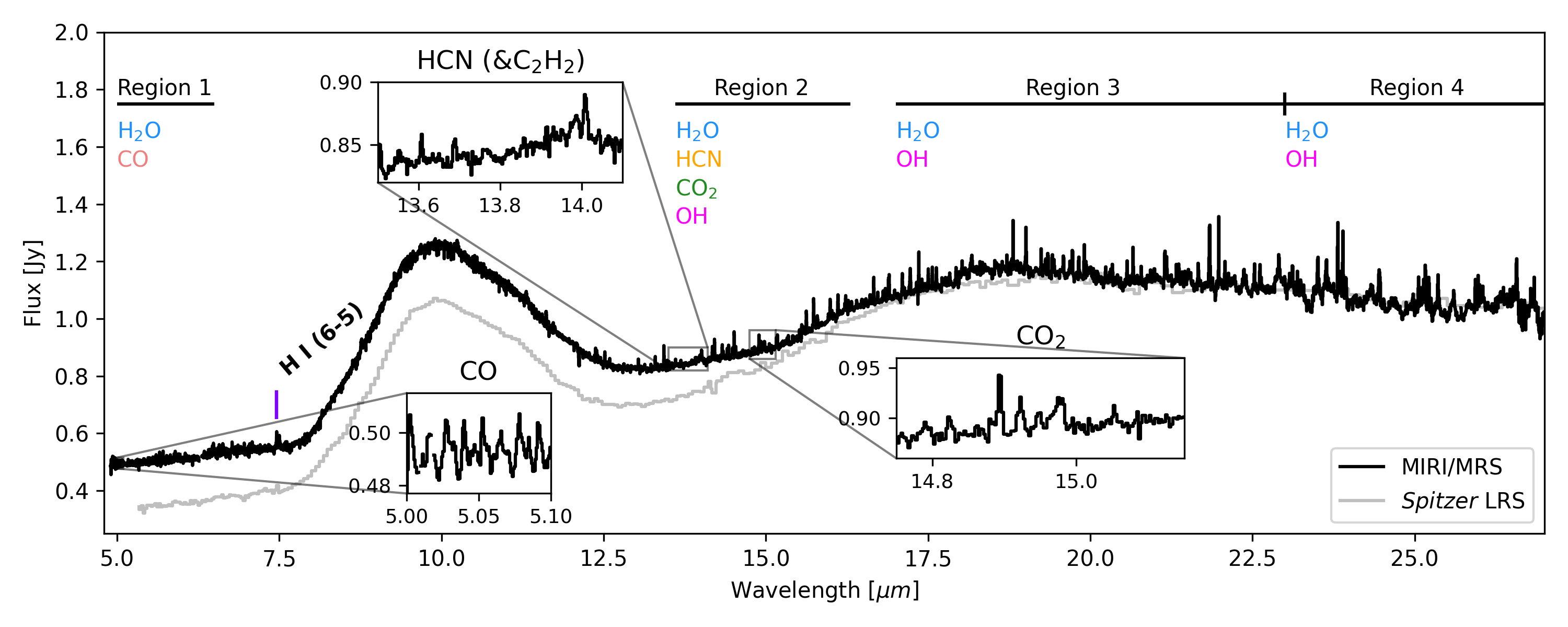}
    \caption{Full MIRI/MRS spectrum of the Sz~98 disk. The extent per region is indicated, along with the detected species. The Spitzer LRS spectrum from the CASSIS database is shown in light grey \citep{ref:11LeBaSp}.}
    \label{fig:full_spec}
\end{figure*}

\section{Methods}
\label{sec:methods}
\subsection{Data acquisition and reduction}
Sz 98 was observed with MIRI/MRS as part of the MIRI mid-INfrared Disk Survey (MINDS) JWST GTO Programme (PID: 1282, PI: T. Henning). Taken on August 8 2022, the exposure of all twelve bands in a four-point dither pattern resulted in an exposure time of approximately 14 minutes per grating setting.

The data were processed using version 1.9.4\footnote{\url{https://jwst-pipeline.readthedocs.io/en/latest/}} of the \textit{JWST} pipeline \citep{ref:22BuEiDe}. We followed the methods described in \citet{ref:23GaArSl} that are specific to point sources to defringe and flux calibrate the data. The reference files exploit the repeatability of the fringes when the pointing is consistent between the science observation and the reference target, and are extracted from the A star HD~163466 (PID: 1050). In the Sz~98 case a more significant pointing offset is seen in channel 2 ($\sim$7.5-11.7~$\mu$m), which is dominated by the silicate feature, but we do not analyse this region in detail. Similarly, a clean spectrophotometric calibration is derived on-sky from HD~163466. This method is adopted mainly for the advantages related to defringing, since the standard pipeline fringe flats require the additional \texttt{residual\_fringe} step to defringe the spectrum. As shown by \citet{ref:23GaArSl}, the latter step may change the shape of molecular features, which would affect the derived excitation properties.

In order to extract the spectrum, the signal in an aperture of 2.5$\times$FWHM centred on the source was summed. The background was estimated from an annulus that linearly grows between 5$\times$FWHM and 7.5$\times$FWHM in the shortest wavelengths, and 3$\times$FWHM to 3.75$\times$FWHM in the longest wavelengths. The aperture correction factors applied are the same as those in \citet{ref:23ArGlLa}, and account for the fraction of the signal of the Point Spread Function (PSF) outside of the aperture, and inside the annulus. Additionally, the outlier detection step of \texttt{spec3} was skipped, since this has spurious results for the data taken with the MRS due to undersampling of the PSF. No rescaling of sub-bands was required to stitch the spectrum.

\subsection{Slab models and fitting procedure}
We subtracted the continuum from the spectrum by fitting a cubic spline through the line-free sections, in order to fit local thermodynamic equilibrium (LTE) slab models. Line-free sections were selected iteratively from visual inspection of the slab fits. Especially in the region of 24~$\mu$m and beyond, the data become noisy and filled with artefacts caused by the low signal in the reference A star HD~163466. Using this knowledge, the continuum in the $>$24~$\mu$m region was selected in such a way that artefacts are avoided. To do so, the continuum points were placed to make the continuum follow along with the larger artefacts where the signal of the A~star drops close to 0 (see \citet{ref:23GaArSl} for more details on how the reference files were derived). Due to the presence of these artefacts, the line fluxes in this region may have a larger uncertainty. However, since we do not analyse this region in great detail due to fewer features of interest, this does not affect our results.

For the \ce{H2O} lines longwards of $\sim$10~$\mu$m, LTE is a good first estimate, though this assumption is less accurate for the lines around 6.5~$\mu$m \citep[e.g.][]{ref:09MePoBl,ref:22BoBeCaa}. The strongest deviation from LTE will be seen in the high-energy lines \citep[$E_{up}>3000$~K, e.g.][]{ref:09MePoBl}. The line profiles were assumed to be Gaussian, with a broadening of $\Delta V=4.7$~km~s$^{-1}$ ($\sigma = 2$~km~s$^{-1}$), similarly to \citet{ref:11SaPoBl}. For molecules with densely packed lines, in other words the \ce{CO2} and \ce{HCN} $Q$-branches, we included mutual shielding from adjacent lines as described in \citet{ref:23TaBeDi}. Subsequently, three parameters were varied in order to fit the features in the observation: the column density $N$, the excitation temperature $T$, and the emitting area $\pi R^2$. The latter scales the strength of the features to match the strength in the observed spectrum. We note that the excitation temperature does not need to be the same as the kinetic temperature of the gas.

The slab models were convolved to a constant resolution per region in the same order as that of the MRS in the relevant bands (ranging from a resolving power $\lambda / \Delta \lambda \approx 3500$ at shorter wavelengths to $\lambda / \Delta \lambda \approx 1500$ at the longer wavelengths, specific values given in Table \ref{tab:chi_results} \citealt{ref:23JoAlSl}), and resampled with \texttt{spectres} \citep{ref:17Ca} to the same wavelength grid. Since we examined the \ce{H2O} features over a large range of wavelengths, the resolution greatly varied between different fits. Additionally, the regions of the disk probed by the different wavelengths and their excitation conditions are not expected to be constant throughout, hence we treated the shorter wavelengths separately from the longer wavelengths \citep[e.g.][]{ref:23BaPoPe}. By dividing the spectrum into the 5-6.5~$\mu$m region, the $\sim$13.6-16.3~$\mu$m region, the $\sim$17-23~$\mu$m region, and 23~$\mu$m and onwards, regions of similar excitation conditions were addressed separately.

Using $\chi^2$ fitting, the most likely values of the variables are found. Similarly to \citet{ref:23GrDiTa}, we first identified relatively bright and isolated lines, used these to find the best fitting slab model, subtracted the model, and moved on to other molecular features. Due to the spectra being very dense in lines, this became an iterative procedure where the noise estimates were taken from the line-subtracted spectra. The order in which we fit the molecules per region is as follows: region 1 - \ce{CO}, \ce{H2O}; region 2 - \ce{H2O}, \ce{OH}, \ce{HCN}, \ce{CO2}; region 3 - \ce{H2O}, \ce{OH}; region 4 - \ce{H2O}, \ce{OH}. The emitting area is parametrised in terms of an arbitrary disk radius $R$. We note that this does not need to be the radius at which the emission is located, but rather the equivalent emitting area. For example, the emission could be confined within a ring of a total emitting area of $\pi R^2$. The number of molecules $\mathcal{N}$ is also included, resulting from the column density and emitting area. It is a more robust metric for optically thin species.

Similarly to \citet{ref:23GrDiTa}, the reduced $\chi^2$ is then defined using:
\begin{equation}
    \chi^2 = \frac{1}{M} \sum^M_{i=1} \frac{F_{obs,i} - F_{mod,i}}{\sigma^2} \text{,}
\end{equation}
where $M$ is the number of data points in the selected wavelength window, $\sigma$ the noise estimated from the selected region with the lines removed, and $F_{obs}$ and $F_{mod}$ are the observed and modelled continuum-subtracted flux, respectively. The confidence intervals are defined as $\chi^2_{min} + 2.3$, $\chi^2_{min} + 6.2$, and $\chi^2_{min} + 11.8$; for $1\sigma$, $2\sigma$, and $3\sigma$, respectively \citep{ref:76Av,ref:92PrTeVe}. The $\sigma$ per region was estimated from a standard deviation on the spectrum itself, after subtracting the best-fit slab models. This was done since, although the visually most `line-free' regions were selected, faint lines were often still present. The resulting spectra and the noise regions are given in Sect. \ref{sec:results}, along with the regions used to fit the lines.

The spacing of the grid is consistent between molecules ($\Delta T=25$~K, $\Delta \log_{10}{(N)}=0.16$~cm$^{-2}$, and $\Delta \log_{10}{(R)}=0.02$~au), but the range was varied depending on how hot or cold the excitation is expected to be. For \ce{OH} much higher temperatures (500-4000 K) were used compared to \ce{CO2} (100-1400 K) and \ce{H2O} (100-1500~K).

\section{Results}
\label{sec:results}

We present the full spectrum in Fig. \ref{fig:full_spec}, where we indicate the molecules detected per region. The overall shape of the continuum is typical of a T~Tauri disk: a discernible silicate feature around 10~$\mu$m indicating the presence of small silicate grains, and excess emission at the longer wavelengths. The silicate feature around 10~$\mu$m is sensitive to changes in grain size, where larger grain sizes indicate more evolved dust with a lower opacity \citep[e.g.][]{ref:01BoMeKo,ref:03PrBoAb,ref:06KeAuDu,ref:10JuBoHe}. The peak value normalised to the continuum is $\sim$1.9, which indicates a relatively small grain size of a few $\mu$m based on the models in \citet{ref:06KeAuDu}. The peak value seems to be slightly below the peak value of EX~Lup, which has since been observed with MIRI/MRS as well \citep{ref:23KoAbDi}.

On top of the continuum, we detect \ce{CO} around 5~$\mu$m, \ce{CO2}, \ce{HCN}, \ce{OH} from $\sim$13~$\mu$m and onwards; and, most strikingly, we see \ce{H2O} features from 5~$\mu$m up to 27~$\mu$m. The relative strength of emission for \ce{CO2} and \ce{H2O} is opposite to the GW~Lup case \citep{ref:23GrDiTa}, where \ce{CO2} was found to be much stronger than \ce{H2O}. Additionally, there is a general lack of detectable carbon-bearing species in the inner disk: aside from \ce{CO2} and \ce{HCN}, we detect no organic molecules. Most notably, \ce{C2H2} is not detected.

\begin{figure*}
    \centering
    \includegraphics[width=\textwidth]{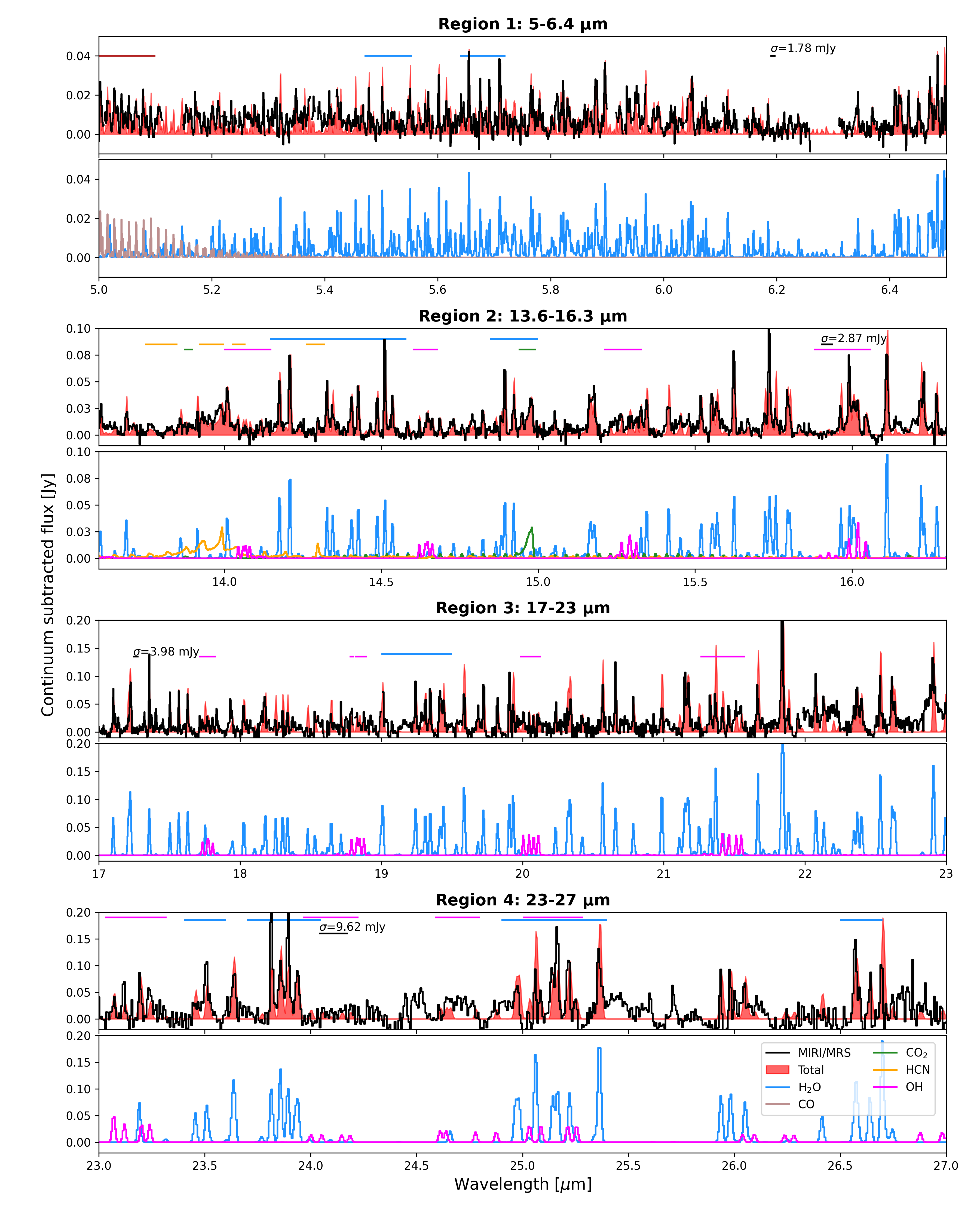}
    \caption{Slab fits for the four different wavelength ranges. The top and bottom panels per region are the data and total model, and the individual slab models; respectively. Spurious spikes and features from the data reduction have been blanked out (see the list in App. \ref{app:blank}). The horizontal lines indicate the regions used to fit the molecules, or the region where $\sigma$ is estimated after subtracting the slab models.}
    \label{fig:spec_ranges}
\end{figure*}

The best-fit parameters are presented in Table \ref{tab:chi_results}, and the corresponding fits per region overlaid on the data can be found in Fig. \ref{fig:spec_ranges}. The $\sigma$ per region are documented in both Table \ref{tab:chi_results} and Fig. \ref{fig:spec_ranges}. The $\chi^2$ maps representing the confidence per fit are included in App. \ref{app:chi2}. Some sections of the continuum subtracted spectrum are negative, particularly in spectral region 4. This is due to the selection of the continuum, which is greatly influenced by the noise and artefacts in this part of the spectrum. We note an up to $\sim 30 \%$ flux discrepancy between the \textit{Spitzer}/IRS and MRS continuum shortwards of $\sim$16~$\mu$m. \citet{ref:17AlMaNa} found a similar discrepancy between photometry data and X-shooter spectroscopy (taken in 2015) of the object, and noted that this was within the expected variability range for Class II young stellar objects found by \citet{ref:14VeBoFl,ref:22FiHiHe}. Additionally, \citet{ref:20BrShGa} classified Sz~98 as a `dipper' star, which typically thought to be caused by the disk being close to edge-on \citep[e.g.][]{ref:15StCoMc,ref:17BoQuAn}. However, based on ALMA millimetre emission, its inclination is 47.1$\degree$ (see \citealt{ref:17TaTeNa} and App. \ref{app:alma}). It is therefore possible that the inner and outer disk are misaligned, causing the discrepancy to be different in the shorter and longer wavelengths.

Despite being located in similar spectral regions, a wide variety of best fit parameters is found for different species. Due to this, we note that it is unlikely they all emit from the same region, as some species may be located deeper or farther out in the disk, and different energy levels are probed per species. Furthermore, the actual spectral overlap is minimal, despite what the MRS resolution might imply. A simple addition of the different models per region is therefore a good approximation. Additionally, simple LTE excitation may not be the most fitting assumption for some species. In the following sections we discuss the best-fit results for the detected molecules in more detail, and some non-detections.

\begin{table}[h]
\centering
\renewcommand{\arraystretch}{1.2}
\caption{\ce{H2O} best-fit model parameters per wavelength region. The number of molecules $\mathcal{N}$ and the average upper energy levels $E_{up}$ per wavelength region are included as well. The confidence intervals are included for for maps with a closed 1$\sigma$-contour (see App. \ref{app:chi2} for more details). The confidence intervals per fit parameter are given based on the $\chi^2$ maps. For $N$, this is given in log-space.}
\label{tab:chi_results}
\resizebox{\columnwidth}{!}{%
\begin{tabular}{cccccc}
\hline \hline
Species                   & \begin{tabular}[c]{@{}c@{}}$N$\\ {[}cm$^{-2}${]}\end{tabular} & \begin{tabular}[c]{@{}c@{}}$T$\\ {[}K{]}\end{tabular} & \begin{tabular}[c]{@{}c@{}}$R$\\ {[}au{]}\end{tabular} &  \begin{tabular}[c]{@{}c@{}}$\mathcal{N}$\\ {[}-{]}\end{tabular} & \begin{tabular}[c]{@{}c@{}}$\overline{E}_{up}$\\ {[}K{]}\end{tabular}\\ \hline
\multicolumn{6}{c}{5-6.5 $\mu$m ($R\sim 3500~\lambda/\Delta \lambda$, $\sigma=1.78$~mJy)}                                                                                                                                                                           \\ \hline
\ce{CO}  & 1.4$\times$10$^{15}$                                        & 1675                                                   & 1.00                         & 9.8$\times$10$^{41}$          &    $\sim$13000            \\
\ce{H2O} & 3.7$\times$10$^{18}$ $_{-1.5}^{+3.1}$                                         & 950 $_{-620}^{+513}$                                                   & 0.07 $_{-0.03}^{+0.46}$                         & -        &     $\sim$7200        \\
\hline
\multicolumn{6}{c}{13.6-16.3 $\mu$m ($R\sim 2500~\lambda/\Delta \lambda$, $\sigma=2.87$~mJy)}                                                                                                                                                                           \\ \hline
\ce{H2O} & 7.9$\times$10$^{18}$ $_{-0.5}^{+0.8}$                                         & 650 $_{-91}^{+88}$                                                   & 0.28 $_{-0.04}^{+0.07}$                         & -       &      $\sim$6500         \\
\ce{CO2} & 2.4$\times$10$^{19}$                                          & 125                                                   & 1.63                          & 4.6$\times$10$^{46}$  & $\sim$3800            \\
\ce{OH}  & 3.6$\times$10$^{13}$                                          & 3075                                                  & 1.87                          & 8.8$\times$10$^{40}$        &   $\sim$13300     \\
\ce{HCN} & 4.3$\times$10$^{16}$                                            & 1075                                                   & 0.76                       & 2.2$\times$10$^{42}$     &   $\sim$6700   \\ \hline
\multicolumn{6}{c}{17-23 $\mu$m ($R\sim 1500~\lambda/\Delta \lambda$, $\sigma=3.98$~mJy)}                                                                                                                                                                           \\ \hline
\ce{H2O} & 7.5$\times$10$^{19}$ $_{-4.2}^{+1.1}$                                       & 300 $_{-41}^{+182}$                                                  & 1.00 $_{-0.39}^{+8.33}$                         & -     &  $\sim$6100    \\
\ce{OH}  & 1.7$\times$10$^{17}$                                         & 950                                                  & 0.28                          & 9.6$\times$10$^{42}$   & $\sim$9600  \\ \hline
\multicolumn{6}{c}{23 $\mu$m onwards ($R\sim 1500~\lambda/\Delta \lambda$, $\sigma=9.62$~mJy)}                                                                                                                                                                      \\ \hline
\ce{H2O} & 1.7$\times$10$^{19}$ $_{-2.0}^{+0.1}$                                        & 250 $_{-44}^{+147}$                                                   & 1.42 $_{-0.66}^{+2.9}$                           & -    & $\sim$6000  \\
\ce{OH}  & \multicolumn{4}{c}{-detected-}  &  $\sim$7700 \\ \hline                                                  
\end{tabular}%
}
\vspace{1ex}
{\raggedright \textbf{Note.} The \ce{OH} fits are poorly constrained, and the best-fit parameters are likely not representative. \par}
\end{table}

\subsection{H$_2$O}
\label{subsec:water}
\begin{figure}[ht!]
    \centering
    \includegraphics[width=\columnwidth]{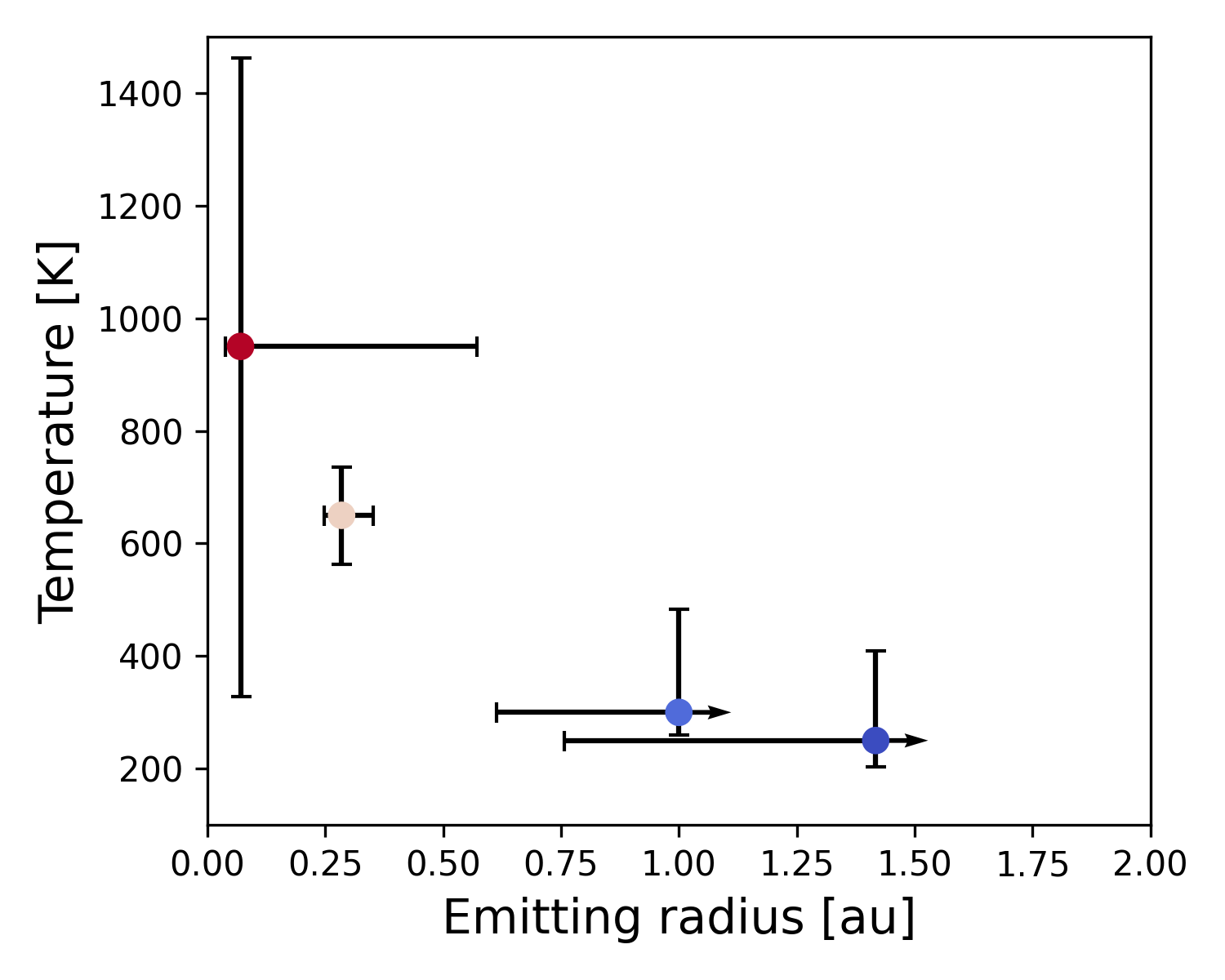}
    \caption{Best-fit temperature and emitting radius for \ce{H2O} per region of the inner disk. The error bars indicate the $1\sigma$ contours of the $\chi^2$ maps.}
    \label{fig:water_radii}
\end{figure}

In Fig. \ref{fig:spec_ranges} we present sections of the continuum-subtracted spectrum where pronounced \ce{H2O} features are present, with the best-fit results of different species. As noted in several previous works \citep[e.g.][]{ref:16BlPoBa,ref:23BaPoPe}, it is expected that the inner (warmer) to outer (colder) regions of the inner disk are probed from shorter to longer wavelengths, respectively. Indeed, we can conclude that this is the case for the spectrum of Sz~98, based on the best-fit parameters presented in Table \ref{tab:chi_results}, and the confidence intervals shown in Fig. \ref{fig:chi2_h2o}. The best-fit temperature and emitting radius change as we move to longer wavelengths. The temperatures slowly decrease from 950 to 250~K, indicating that we are gradually probing colder and/or less excited gas. We present this tentative trend in Fig. \ref{fig:water_radii}. The error bars are based on the $1\sigma$-contours of the $\chi^2$ plots in Fig. \ref{fig:chi2_h2o}. The best-fit slab model of region 1 will underestimate the lines in region 4, and vice versa. Furthermore, adding the \ce{H2O} spectra from all regions together significantly overestimates the flux. In reality a specific disk region of a certain temperature and radial extent will not be contained to a specific spectral region, but influences lines in other regions as well. Alternatively, adding the spectra together by assuming the radii found are instead the inner and outer radii of a series of annuli, the spectrum is better reproduced while keeping the radii within the $1\sigma$-confidence intervals. This indicates that one slab cannot be used to fit the entire MIRI/MRS range, and future work must accommodate for a temperature and emitting area gradient in the fitting procedure.

Additionally, we systematically find the \ce{H2O} lines to be optically thick, and our \ce{H2O} column densities are in a similar range as those found previously from \textit{Spitzer} spectra of other disks \citep{ref:11CaNa,ref:11SaPoBl}, although in regions 3 and 4 they are higher than usually inferred. As demonstrated by the models of \citet{ref:09MePoBl,ref:15WaNoDi}, we are likely not probing the full column density of \ce{H2O} down to the mid-plane, especially in the short wavelength region. Part of the \ce{H2O} gas is hidden below the dust continuum where $\tau_{IR}\approx 1$, where dust might be blocking emission from the deeper layers in the disk. The small emitting radius of \ce{H2O} in spectral region 1 indicates that we would be probing the inner gas disk \citep[e.g.][]{ref:10DuMo}.

Not all \ce{H2O} lines are fit equally well. \citet{ref:09MePoBl} show that high energy lines are less likely to be thermally excited, and might be better fit with lower temperature slab models. This would be a sign that some lines are sub-thermally excited, and the LTE assumption is not applicable. However, the LTE slab models fit the spectrum very well, therefore no significant evidence for non-LTE excitation of \ce{H2O} is found here. While more detailed thermo-chemical models could result in a better representation of the spectrum, the increased complexity introduces more uncertainties in the fits, and this is left for future work, along with fits of temperature gradients. However, the temperature trend seen in the simple slab models of the different regions is robust.

\begin{figure}[ht!]
    \centering
    \includegraphics[width=0.94\columnwidth]{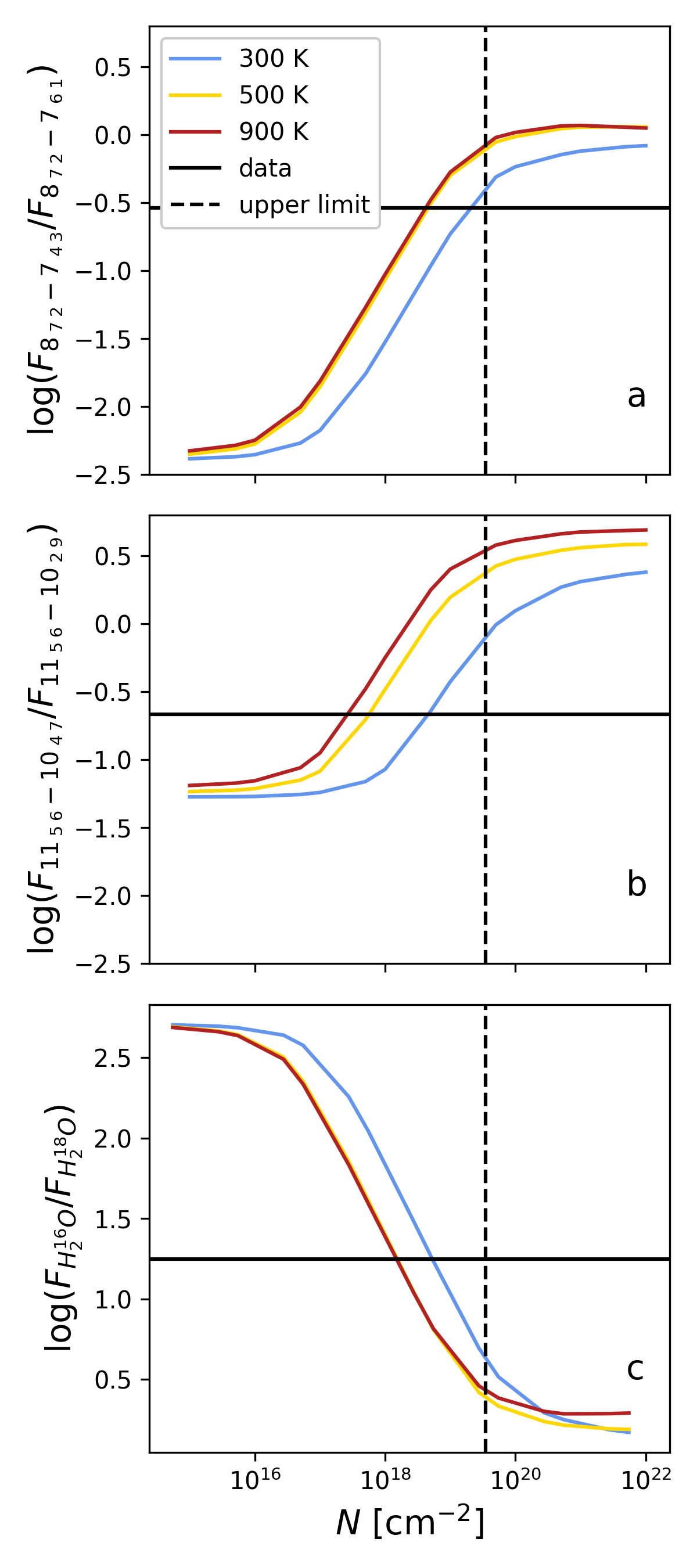}
    \caption{Flux ratios of H$_\text{2}^{\text{16}}$O lines of the same upper level in the data compared to slab models (top); and H$_\text{2}^{\text{16}}$O/H$_\text{2}^{\text{18}}$O ratio in the data and slab models (bottom). The properties of the transitions can be found in Table \ref{tab:opacity}. The dotted black vertical line indicates the upper limit of the column density based on the $1\sigma$ level of the spectrum and the non-detection of the H$_\text{2}^{\text{18}}$O line. The x-axis shows the column density of H$_\text{2}^{\text{16}}$O. In the bottom panel, H$_\text{2}^{\text{16}}$O/H$_\text{2}^{\text{18}}$O ratio of 550 is assumed.}
    \label{fig:water_opacity}
\end{figure}

In order to assess the correctness of the $\chi^2$ fit results, we compare the fluxes of pairs of lines with the same upper energy levels, but different Einstein $A_{ul}$ coefficients. The flux ratio of these lines will depend primarily on the opacity of the lines rather than the temperature, providing a robust estimate for the column density. To reduce the effects of artefacts and other molecular features, we limit the examined range to isolated lines. In this manner, two pairs of lines were identified that are primarily sensitive to the changes in column density. The properties of these lines can be found in Table \ref{tab:opacity}. Evaluating the ratios of these lines in slab models of changing column density and temperature results in the coloured trends in panels \textbf{a} and \textbf{b} of Fig. \ref{fig:water_opacity}, while the black horizontal lines result from the flux ratios in the data. For low column densities below $\sim$10$^{16}$~cm$^{-2}$-10$^{18}$~cm$^{-2}$ (depending on the temperature assumed), the trends are largely flat, since the flux ratio depends on the $A_{ul}$ ratio. Once one of the lines becomes opaque, this ratio will change, resulting in the upward trend for higher column densities. When assessing the observed flux ratios in the data the column density should indeed be high as suggested by the best-fit slab models. For optically thin lines, the flux ratios converge to a single value that corresponds to the $A_{ul}$ ratio. This is not the case, as shown in Table \ref{tab:opacity}, indicating that the brighter line could be opaque. However, we note that these lines are part of a cluster of lines, and the ratios are likely affected by line blending. In that case, we cannot claim that the lines are optically thick based on the fact that the ratio of $A_{ul}$ does not equal the flux ratio. However, the flux ratios of the slab models presented in Fig. \ref{fig:water_opacity} are similarly affected, therefore these trends are still representative.

Note that it is assumed that the emitting area of the line pairs is equal. Assuming the longer wavelength line traces a larger area, the larger the discrepancy between the emitting areas, the higher the flux of the reference line at longer wavelengths, and the more the trends are shifted down. Therefore, even for differing emitting areas, this conclusion remains valid.

\begin{table*}[h]
\centering
\caption{Line pairs used to assess the opacity. The reference lines are the lines with the highest $A_{ul}$. The corresponding panel (\textbf{a}, \textbf{b} or \textbf{c}) is indicated.}
\label{tab:opacity}
\resizebox{0.8\textwidth}{!}{%
\begin{tabular}{lllll}
\multicolumn{1}{c}{\begin{tabular}[c]{@{}c@{}}Wavelength\\ {[}$\mu$m{]}\end{tabular}} & \begin{tabular}[c]{@{}l@{}}Transition (upper-lower levels)\\ Level format: $v_1v_2v_3 \text{ } J_{K_aK_c}$\end{tabular} & \multicolumn{1}{c}{\begin{tabular}[c]{@{}c@{}}$A_{ul}$\\ {[}s$^{-1}${]}\end{tabular}} & \multicolumn{1}{c}{\begin{tabular}[c]{@{}c@{}}$E_u$\\ {[}K{]}\end{tabular}} & \multicolumn{1}{c}{\begin{tabular}[c]{@{}c@{}}$\log((F/F_{max})/(A_{ul}/A_{ul,max}))$\\ {[}-{]}\end{tabular}} \\ \hline \hline
\multicolumn{5}{c}{Line pair \textbf{a}}                                                       \\ \hline
26.70259 (ref)                                                                        & 000-000 8$_{\text{ 7 2}}$ -- 7$_{\text{ 6 1}}$                                                                                & 21.48                                                                             & 2288.63        &    \multicolumn{1}{c}{2.18}                                                      \\
15.16408                                                                              & 000-000 8$_{\text{ 7 2}}$ -- 7$_{\text{ 4 3}}$                                                                                & 0.06                                                                              & 2288.63  &  \\ \hline
\multicolumn{5}{c}{Line pair \textbf{b}}                                                                                                          \\ \hline
23.94296 (ref)                                                                        & 000-000 11$_{\text{ 5 6}}$ -- 10$_{\text{ 4 7}}$                                                                              & 9.98                                                                              & 2876.09      &   \multicolumn{1}{c}{1.23}                                                         \\
14.17713                                                                              & 000-000 11$_{\text{ 5 6}}$ -- 10$_{\text{ 2 9}}$                                                                              & 0.30                                                                              & 2876.09      &       

\\ \hline
\multicolumn{5}{c}{Line pair \textbf{c}}                                                                                                          \\ \hline
26.70220 (H$_\text{2}^{\text{16}}$O)                                                                        & 000-000 8$_{\text{ 7 1}}$ -- 7$_{\text{ 6 2}}$                                                                              & 21.48                                                                              & 2288.65      &   \multicolumn{1}{c}{-}                                                         \\
26.98991 (H$_\text{2}^{\text{18}}$O)  &   000-000 8$_{\text{ 7 1}}$ -- 7$_{\text{ 6 2}}$     & 20.77                                                                              & 2265.60      &       

\end{tabular}%
}
\end{table*}

\begin{figure*}[h]
    \centering
    \includegraphics[width=\textwidth]{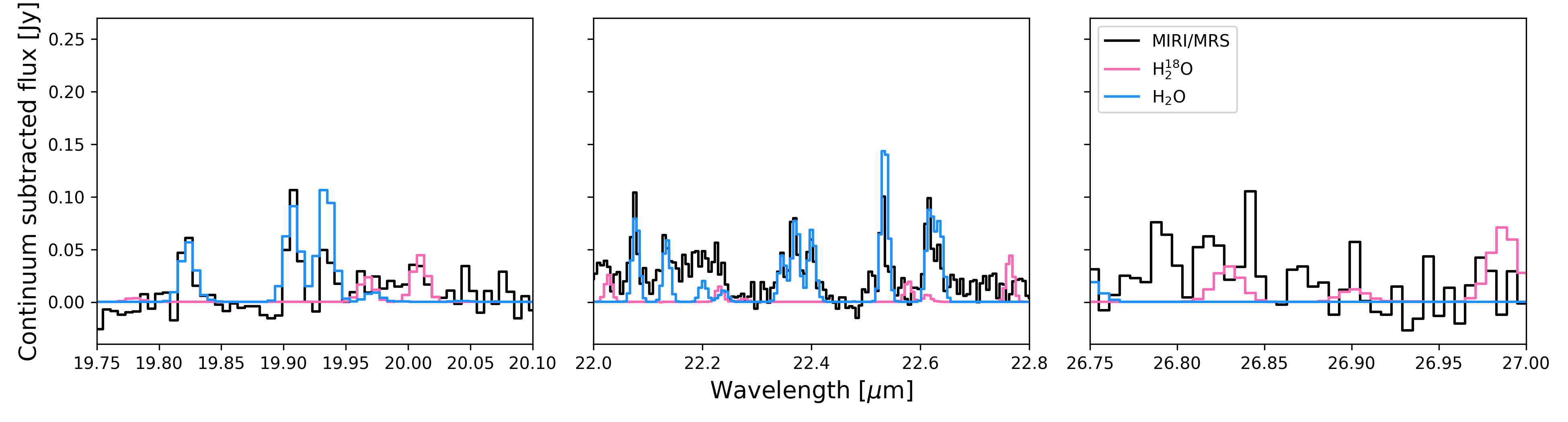}
    \caption{Different regions where H$_\text{2}^{\text{18}}$O is expected to show lines sufficiently isolated from \ce{H2O}. These lines are not seen in the spectrum, compared a H$_\text{2}^{\text{18}}$O slab models assuming the same best-fit parameters as \ce{H2O} in these regions, assuming a $^{16}O/^{18}O$ of 550. The models have been multiplied by 5 for visibility. The \ce{H2O} slab models are based on the best-fit parameters presented in Table \ref{tab:chi_results} per wavelength region.}
    \label{fig:h218o}
\end{figure*}

Since H$_\text{2}^{\text{18}}$O-rich regions are expected to be located deeper in the disk atmosphere emitting optically thin lines, detection and characterisation of the isotopologue would put constraints on the \ce{H2O} column densities, and allow for the abundance to be measured \citep{ref:22CaBeBo}. Several isolated lines are potentially detectable by MIRI/MRS, which we plot in Fig.~\ref{fig:h218o}. Given the large column density of \ce{H2O}, the isolated H$_\text{2}^{\text{18}}$O could potentially be visible. We compare the data to slabs of the same excitation temperature and emitting radius as the best-fit \ce{H2O} values in the wavelength region assuming a $^{16}O/^{18}O$ of 550. In Fig.~\ref{fig:h218o} we demonstrate that H$_\text{2}^{\text{18}}$O is not detected in our data. One of the brightest, isolated lines in the spectrum is the line around 27~$\mu$m, the details of which can be found in Table \ref{tab:opacity}. Since it is not detected, the lower limit on the line flux of the H$_\text{2}^{\text{18}}$O line (taken from 26.9626 to 27.0106~$\mu$m) is $\sim$0.4$\times 10^{-14}$~erg~s$^{-1}$~cm$^{-2}$ normalised to a distance of 140~pc. Assuming the integrated flux of the H$_\text{2}^{\text{18}}$O feature is in this upper limit on the line flux, the maximum column density of \ce{H2O} is 3.5$\times$10$^{19}$~cm$^{-2}$ when assuming the ISM H$_\text{2}^{\text{16}}$O/H$_\text{2}^{\text{18}}$O ratio of 550. This value has been indicated as an upper limit in Fig. \ref{fig:water_opacity}, and in the $\chi^2$ maps in App. \ref{app:chi2}. Based on the non-detection, we can plot a comparison to the slab model line ratios in a similar fashion as for \ce{H2O}. However, now the comparison is between the H$_\text{2}^{\text{18}}$O line and a H$_\text{2}^{\text{16}}$O line of similar upper energy (see Table \ref{tab:opacity} for the details), shown in panel \textbf{c} of Fig. \ref{fig:water_opacity}. Note that the column density on the x-axis and in the slab model of H$_\text{2}^{\text{16}}$O is multiplied by 550 compared to the column density of H$_\text{2}^{\text{18}}$O. Fig. \ref{fig:water_opacity} indicates that the column densities are indeed high, but likely between 5$\times$10$^{18}$ to 3.5$\times$10$^{19}$~cm$^{-2}$ rather than 7.5$\times$10$^{19}$~cm$^{-2}$ as found for region 3 in Table \ref{tab:chi_results}. The larger column density might be possible, provided the H$_\text{2}^{\text{16}}$O/H$_\text{2}^{\text{18}}$O ratio is larger than 550. Some variation in this ratio per transition might be possible \citep[see e.g.][]{ref:22CaBeBo}.

\subsection{CO}
The best-fit parameters for the \ce{CO} emission in the disk are 1675~K, with a radius of 1.00~au, and a column density of 1.4$\times$10$^{15}$~cm$^{-2}$. However, we note a degeneracy between the three parameters, causing the confidence intervals to be relatively wide. The best fit excitation temperature found here is not likely equal to the true gas temperature, which is likely colder than the best fit of 1675~K and optically thick, since the gas temperatures probed by \ce{CO} are typically not more than 1000~K \citep[e.g.][]{ref:13BrPoDi,ref:22BaAbBr,ref:21AnBlCl}. As shown in Table \ref{tab:chi_results}, the upper energy levels in this wavelength region are typically much higher than that of \ce{H2O}; therefore, we are likely not tracing the same region for both species. Only the rotationally excited \ce{CO} lines of the v=1-0 band can be detected by MIRI/MRS. The \ce{CO} lines in the Near Infrared Spectrograph (NIRSpec) region and observed from the ground for bright sources can be expected to be probing more similar conditions.

\subsection{CO$_2$}
We detect the \ce{CO2} $Q$-branch, although the hot bands are not confidently detected, similarly to the isotopologue $^{\text{13}}$CO${\text{2}}$. The \ce{CO2} emission is much weaker compared to the \ce{H2O} emission, which is opposite to the case of GW~Lup \citep{ref:23GrDiTa}. Due to the blending with \ce{H2O} lines and faintness of the features, it is difficult to constrain the excitation properties. A demonstration of the differences between optically thick and optically thin emission with $^{\text{13}}$CO$_\text{2}$ can be found in App. \ref{app:co2}. The best-fit temperature when including the hot band around 13.8~$\mu$m is relatively cold, with 125~K from an emitting radius of 1.63~au, and a column density of $\sim$2.4$\times 10^{19}$~cm$^{-2}$. Due to the lower excitation temperature, it likely originates from a deeper layer in the disk, or farther away from the star. However, when fitting only the $Q$-branch, the best fit parameters are an excitation temperature of 250~K, and emitting radius of 0.28~au (see Fig. \ref{fig:chi2_region2}). While the emitting area in this case is similar to that of \ce{H2O}, the excitation temperature is different.

\subsection{OH}
The maximum $E_u$ of the \ce{OH} transitions we tentatively detect is $\sim$15000~K, corresponding to an upper rotational quantum number of $\sim$25 \citep{ref:21TaHeDi}. The \ce{OH} emission, though poorly constrained, seems to have a high excitation temperature (likely nearer 2000~K or higher) with a lower column density than \ce{H2O} ($<$2$\times$10$^{17}$~cm$^{-2}$) in all regions where it is detectable. Under the LTE assumption, the emission could originate from high in the disk atmosphere, based on these elevated temperatures. However, other explanations are possible. We further discuss the excitation of \ce{OH} in this context in Sect. \ref{subsec:prompt_oh}.

\subsection{HCN}
The broad \ce{HCN} $Q$-branch is detected in spectral region 2, around 14~$\mu$m; along with a hot band around 14.3~$\mu$m. The constraints on the fit are poor due to the degeneracy between the column density and the emitting radius for low column density, and the degeneracy between the column density and temperature for high column density. While \ce{HCN} can be fit with both an optically thin and optically thick solution, we find a best fit for a temperature of 1075~K, with a radius of 0.76~au, and a column density of 4.3$\times$10$^{16}$~cm$^{-2}$. However, based on thermo-chemical models, it is more likely in the range of 330~K at a much higher column density \citep{ref:18WoMiTh}, which is still contained in the $1\sigma$-confidence level (see Fig. \ref{fig:chi2_region2}). This is far higher than previously found from \textit{Spitzer} data \citep[e.g.][]{ref:11SaPoBl,ref:11CaNa}, which could be an incorrect fit resulting from the lower resolution. For example, \citet{ref:23GrDiTa} showed that the column density of \ce{CO2} inferred from MIRI/MRS data must be much higher than previously assumed for GW~Lup.

\subsection{Hydrogen}
We report the detection of a hydrogen recombination line. The strongest hydrogen recombination line, HI~(6-5) at $\sim$7.5~$\mu$m, is notably weaker than in GW~Lup, where [NeII] was also detected \citep{ref:23GrDiTa}. The location of this line is indicated in Fig. \ref{fig:full_spec}. Surprisingly, the HI~(7-6) line, which is thought to trace accretion, is not detected or hidden due to a \ce{H2O} line \citep{ref:15RiPaDu}, despite Sz~98 being an active accretor \citep{ref:08MeJoSp,ref:11MoOlVa}.

\subsection{Non-detections}
Notably, some species are not detected currently. Among these are the hydrocarbons, \ce{C2H2} and \ce{CH4}, of which the former has been detected in GW~Lup \citep{ref:23GrDiTa}. \ce{NH3} is also not present. Despite being a strong accretor, no [NeII] is found. Furthermore, no molecular hydrogen is found to be strong enough to be detected in the forest of \ce{H2O} lines. As mentioned above, while \ce{H2O} is easily visible in the spectrum, its isotopologue H$_\text{2}^{\text{18}}$O is not.

\section{Discussion}
\label{sec:discussion}
As presented in Sect. \ref{sec:results}, one of the defining features of the spectrum of Sz~98 is that \ce{H2O} emission is dominant, compared to relatively weak \ce{CO2} and a lack of \ce{C2H2}, together indicating a \ce{C/O} ratio below unity. This is surprising, considering the large disk size and presence of dust traps, which should indicate limited drift of icy pebbles and thus correlate with a relative increase in carbon-bearing molecules \citep{ref:20BaPaBo}. However, this correlation depends on assumptions related to the ice composition, formation timescale of substructures, among others; and may therefore change with time \citep[e.g.][]{ref:15PiObBi}. In contrast, \citet{ref:19MiFaVa} inferred that the volatile \ce{C/O} ratio is $>$1 in the outer disk of Sz~98 based on bright \ce{C2H} emission, inconsistent with our findings for the gas composition of the inner disk. We discuss a number of processes that may explain the observation of the inner disk of Sz~98.

\begin{figure}[ht!]
    \centering
    \includegraphics[width=0.9\columnwidth]{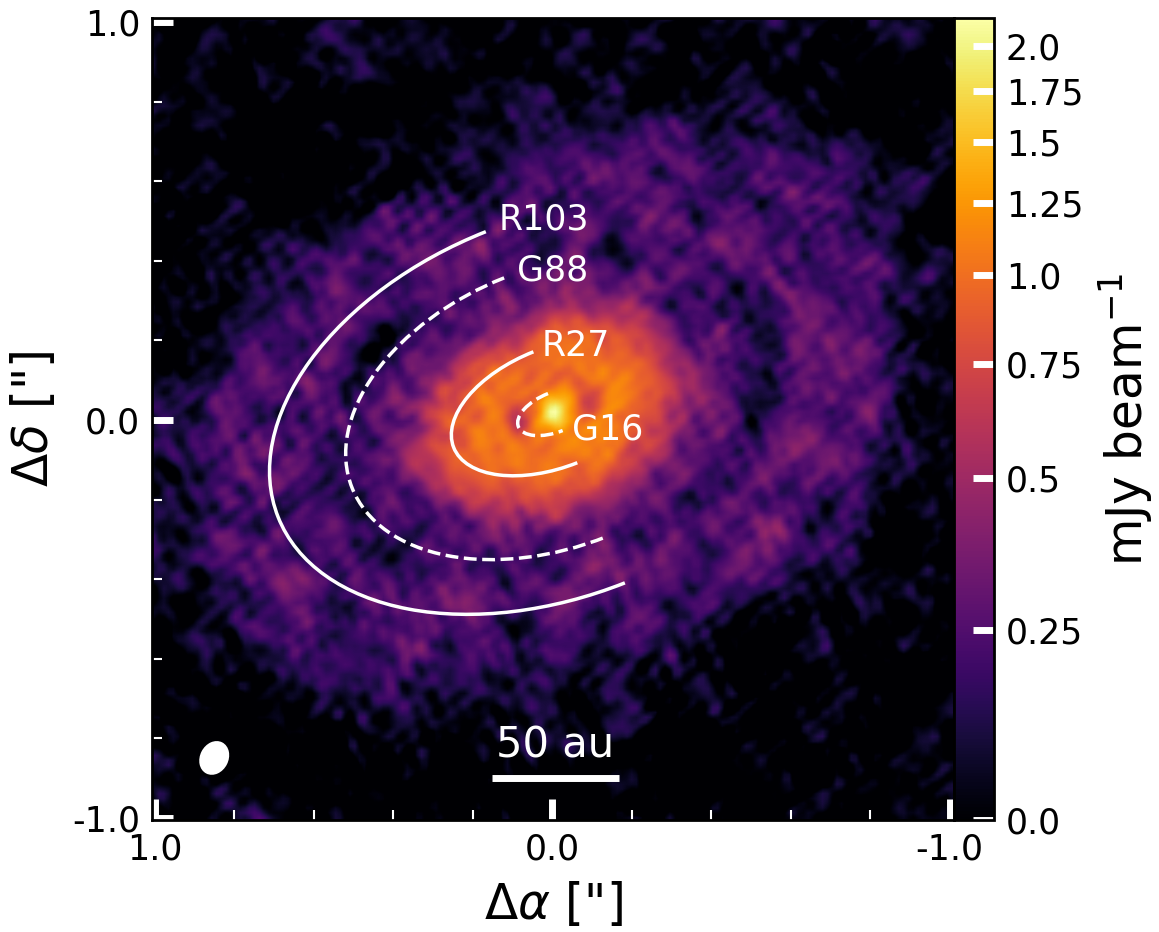}
    \caption{ALMA millimetre dust continuum emission from Sz~98 centred at 232.984 GHz (1286.75~$\mu$m). The dotted lines indicate gaps (G), the solid lines indicate rings (R), and their radii are included in au. Appendix \ref{app:alma} describes how the image is created.}
    \label{fig:continuum}
\end{figure}

\subsection{Radial drift increasing the H$_2$O-gas reservoir}
\label{subsec:rad_drift}
Despite being a very large and extended dust disk, the millimetre continuum shows a considerable central concentration resulting in the characteristic `fried egg' shape found in \citet{ref:18TeDiAn} and presented in Fig.~\ref{fig:continuum}. While various explanations are possible for this characteristic shape, it could be indicative of radial transport inwards. If the bright central continuum is indeed due to pebble drift, this radial transport means pebbles containing \ce{H2O}-ice fall inwards within the \ce{H2O} snow line, estimated to lie at $\sim$1.1 au (see App. \ref{app:alma}). This results in an increase in the volatile oxygen reservoir due to sublimation of \ce{H2O} ice \citep[e.g.][]{ref:06CiCu,ref:20BaPaBo}. The inner disk is expected to be more replenished in oxygen-bearing species compared to carbon due to the snow lines of \ce{CO} and \ce{CO2} being farther out than that of \ce{H2O} \citep[e.g.][]{ref:21ObBe}. The \ce{CO2} and \ce{CO} snow lines are estimated to lie at 2.2~au and 20~au, respectively, in App. \ref{app:alma}. Therefore, despite not conforming with the disk mass or size versus carbon bearing species correlation \citep{ref:11CaNa,ref:13NaCaPo,ref:20BaPaBo}, radial transport could still be the cause of the large \ce{H2O} column in the inner disk. This could indicate that the substructures of the disk are either not 100\% efficient at trapping pebbles \citep[see e.g.][]{ref:22StMcHa}, or were formed late in the disk evolution when the pebbles containing \ce{H2O}-ice had already drifted inwards. The age of Sz~98 has been estimated to be between 1.7 and 5.6~Myr \citep{ref:19VaRuDi}, and typical drifting timescales can be anywhere between $\sim$10~kyr and $\sim$1~Myr starting shortly after disk formation \citep{ref:12BiKlEr,ref:15BiAnPi}. On the other hand, the timescale on which substructures form is uncertain, although relatively young objects exist that already show gaps, such as HL~Tau \citep{ref:15AlBrPe} ($\sim$1~Myr old, \citealt{ref:19VaRuDi}). Therefore, pebbles could have drifted inwards prior to the formation of the gaps, but more information about gap formation is required to confirm this.


\subsection{H$_2$O across the inner disk}
The \ce{H2O} lines in the mid-infrared are thought to probe the $\sim$0.1~au region out to the \ce{H2O} snow line \citep[e.g.][]{ref:17BaPoSa,ref:23BaPoPe}. The critical densities to excite \ce{H2O} emission are higher ($\sim 10^{13}$~cm$^{-3}$) at shorter wavelengths for the vibrational bending modes; and lower ($\sim 10^{8}$~cm$^{-3}$) for the rotational lines at the longer wavelengths \citep{ref:09MePoBl}. This indicates that excitation due to collisions is less efficient for the vibrational bending modes, requiring a larger density for thermal excitation. Similarly, the rotational lines are more easily excited, requiring lower density. Indeed, in the shorter wavelengths around 3-4~$\mu$m, previous studies using a variety of ground-based instruments have found high temperature (1500~K), optically thick (10$^{20}$~cm$^{-2}$) \ce{H2O} emission in other disks \citep{ref:04CaToNa,ref:09SaBlBo,ref:11DoNaCa,ref:22SaPoBa,ref:23BaPoPe}. Moving outwards to the longer wavelengths probed by the MRS, the emission is well fit with 300-600~$K$ with a column density of 10$^{18}$~cm$^{-2}$ based on \textit{Spitzer} IRS \citep[e.g.][]{ref:11CaNa,ref:11SaPoBl}.

These ranges are not unlike what we find here (see Table \ref{tab:chi_results}): the $\sim 5$~$\mu$m range shows hotter emission around 950~K, while the best-fit temperatures slowly decrease down to 250-650~K at the longer wavelengths; although we find higher column densities in regions 3 and 4. Additionally, the column density also shows the opposite trend, being higher for regions 2 and 3 compared to region 1. However, we note that the maps in Fig. \ref{fig:chi2_h2o} show that the fits are less well-constrained in terms of column density, and could all very well be in the 10$^{18}$~cm$^{-2}$ range still. Alternatively, we could be probing a change in dust opacity \citep{ref:15AnKaRi}, where $\tau_{dust}$ is located higher up in the disk closer to the star. Additionally, \citet{ref:23BaPoPe} find the slab model fits to result in similar column densities for different sources throughout different wavelengths, of the order of $\sim$10$^{18}$~cm$^{-2}$. They therefore suggest that the different wavelengths probe the conditions where excitation is met per disk radius, resulting in decreasing temperatures and increasing emitting radii, at similar column densities. Since the column density is not well-constrained, this may still be true here. 

Finally, some disks in the sample of \citet{ref:10PoSaBl} exhibit relatively lower line fluxes at longer wavelengths in \textit{Spitzer}, pointing to some depletion past the mid-plane snow line, due to the `cold finger effect', which results in vertical transport across the snow line. In App. \ref{app:alma}, the snow line is found to be located at 1.1~au. Due to the brightness of the lines in region 4, it is unlikely that a significant amount of \ce{H2O} is transported down to the mid-plane across the \ce{H2O} snow line in Sz~98, since this would result in relatively weaker lines in region 4 due to formation of \ce{H2O}-ice \citep[see also][]{ref:10PoSaBl}. This indicates that a high abundance of \ce{H2O} is present across the disk surface. Furthermore, if the emitting radius of 1.4~au for region 4 found in Table \ref{tab:chi_results} corresponds to the true emitting radius, it is indeed close to the 1.1~au snow line. However, far-IR \ce{H2O} past the MRS wavelength range would be more telling regarding the presence or absence of a `cold finger effect', as done in \citet{ref:16BlPoBa}.


\begin{figure}
    \centering
    \includegraphics[width=\columnwidth]{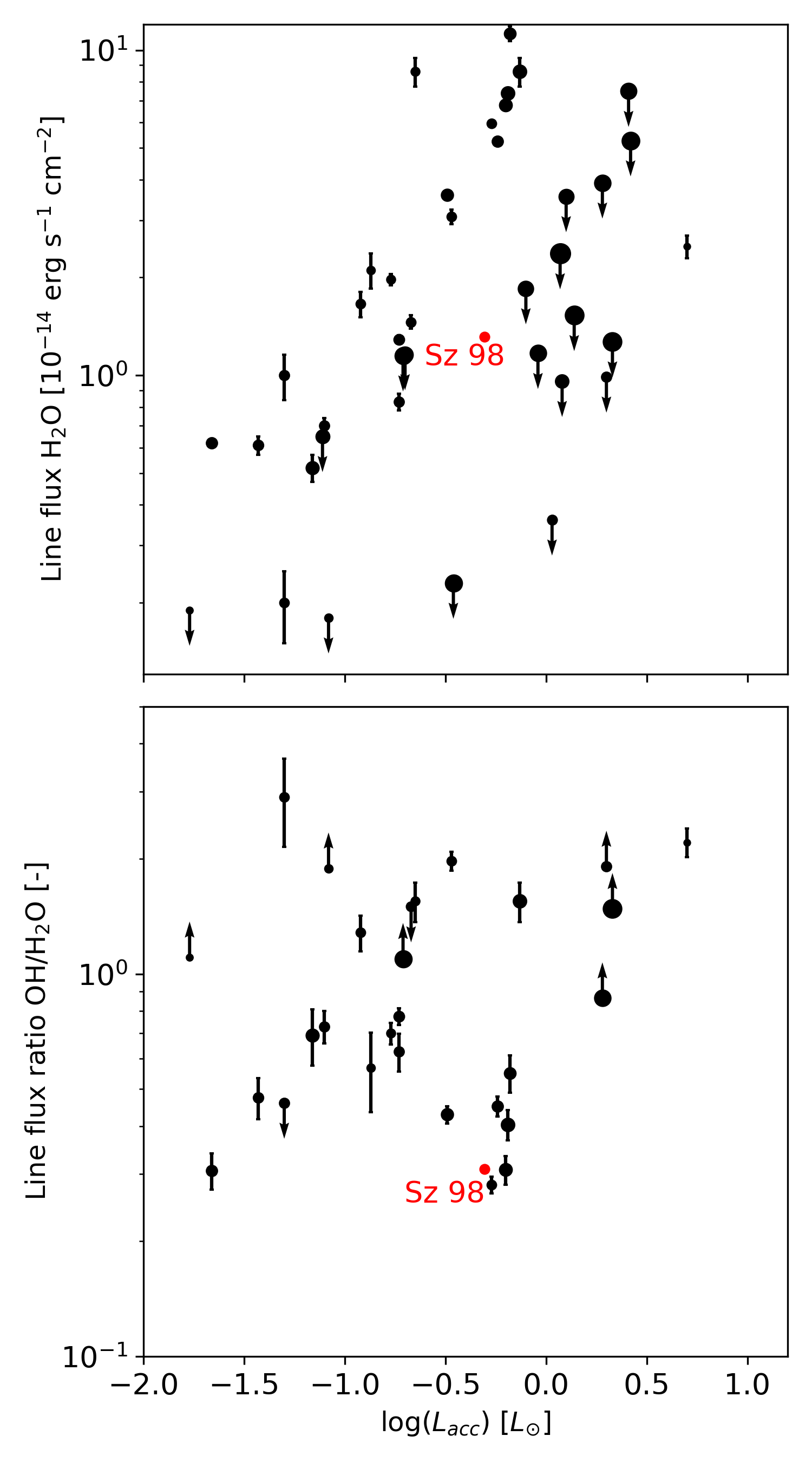}
    \caption{Line fluxes of \ce{H2O} (top) and line flux ratio \ce{OH}/\ce{H2O} (bottom) and accretion luminosities from the disk sample in \citet{ref:17BaPoSa} compared to Sz~98 (red dot). The size of the circle is proportional to the stellar mass, and the arrows indicate upper and lower limits. All fluxes have been normalised to 140~pc.}
    \label{fig:line_fluxes}
\end{figure}

\subsection{H$_2$O self-shielding}
\ce{H2O} is capable of self-shielding against UV radiation when it has sufficiently high column densities, starting at $\sim$2$\times$10$^{17}$~cm$^{-2}$ \citep{ref:09BeBe,ref:17HeBoDi}. The column densities inferred here (see Table \ref{tab:chi_results}), are well above this range. In this case, UV radiation cannot penetrate deep into the \ce{H2O} column, resulting in less photodissociation of \ce{H2O}, and potentially shielding other species \citep{ref:22BoBeCaa}.

As mentioned above, \ce{OH} can be recycled back from \ce{H2O} through photodissociation of \ce{H2O}. Therefore, \ce{H2O} self-shielding results in a decreased \ce{OH} abundance. The upper disk will be exposed to irradiation, resulting in a relatively thin \ce{OH} layer above the self-shielding \ce{H2O} column \citep{ref:15WaNoDi}. Therefore, the gas containing \ce{OH} is expected to be hot, with a lower column density. Assuming the excitation is in LTE, the fact that high energy lines are clearly observed at least in the $\sim$13--16~$\mu$m range indicates that the \ce{OH} gas must be hot. Indeed, from our fit (see Table \ref{tab:chi_results}) this is what we find. However, both prompt emission and chemical pumping from \ce{O + H2} can also excite these higher energy lines. We discuss this further in Sect. \ref{subsec:prompt_oh}

\subsection{H$_2$O and OH line fluxes}
The \ce{OH} emission is relatively weak in Fig.~\ref{fig:spec_ranges} compared to \ce{H2O}. \citet{ref:17BaPoSa} examine the line fluxes of the 12.52~$\mu$m and the tentatively detected 12.6~$\mu$m \ce{H2O} and \ce{OH} lines, respectively, of a sample of disks observed with \textit{Spitzer}. We plot the same line fluxes from these \textit{Spitzer}/IRS spectra compared to the accretion luminosity in Fig. \ref{fig:line_fluxes}, where in the top panel only the \ce{H2O} line flux is shown, and the bottom panel the \ce{OH}/\ce{H2O}-ratio. For samples where only an upper limit can be given for both \ce{H2O} and \ce{OH}, the ratio is not included. The line fluxes for Sz~98 are calculated from the MIRI/MRS spectrum over the same interval, with the best-fit slab models of other species subtracted. Generally, the line flux of \ce{H2O} increases with stellar luminosity \citep{ref:11SaPoBl}, and stellar mass. From the top panel of Fig. \ref{fig:line_fluxes}, it becomes clear that the flux of the 12.52~$\mu$m line is on the lower end for Sz~98, but not out of the ordinary. A lower line flux may indicate that the inner disk has a comparatively higher amount of small dust blocking part of the \ce{H2O} column \citep{ref:17AnBrKa}. In this case part of the radiation can be extincted by the dust, rather than the \ce{H2O} itself, making \ce{H2O} self-shielding less important for the chemistry in the inner disk. Additionally, the line flux ratio with \ce{OH} shows a similar trend. The line flux of \ce{OH} compared to \ce{H2O} is relatively low, but not unique in the sample.

\subsection{OH prompt emission}
\label{subsec:prompt_oh}
\begin{figure*}
    \centering
    \includegraphics[width=0.9\textwidth]{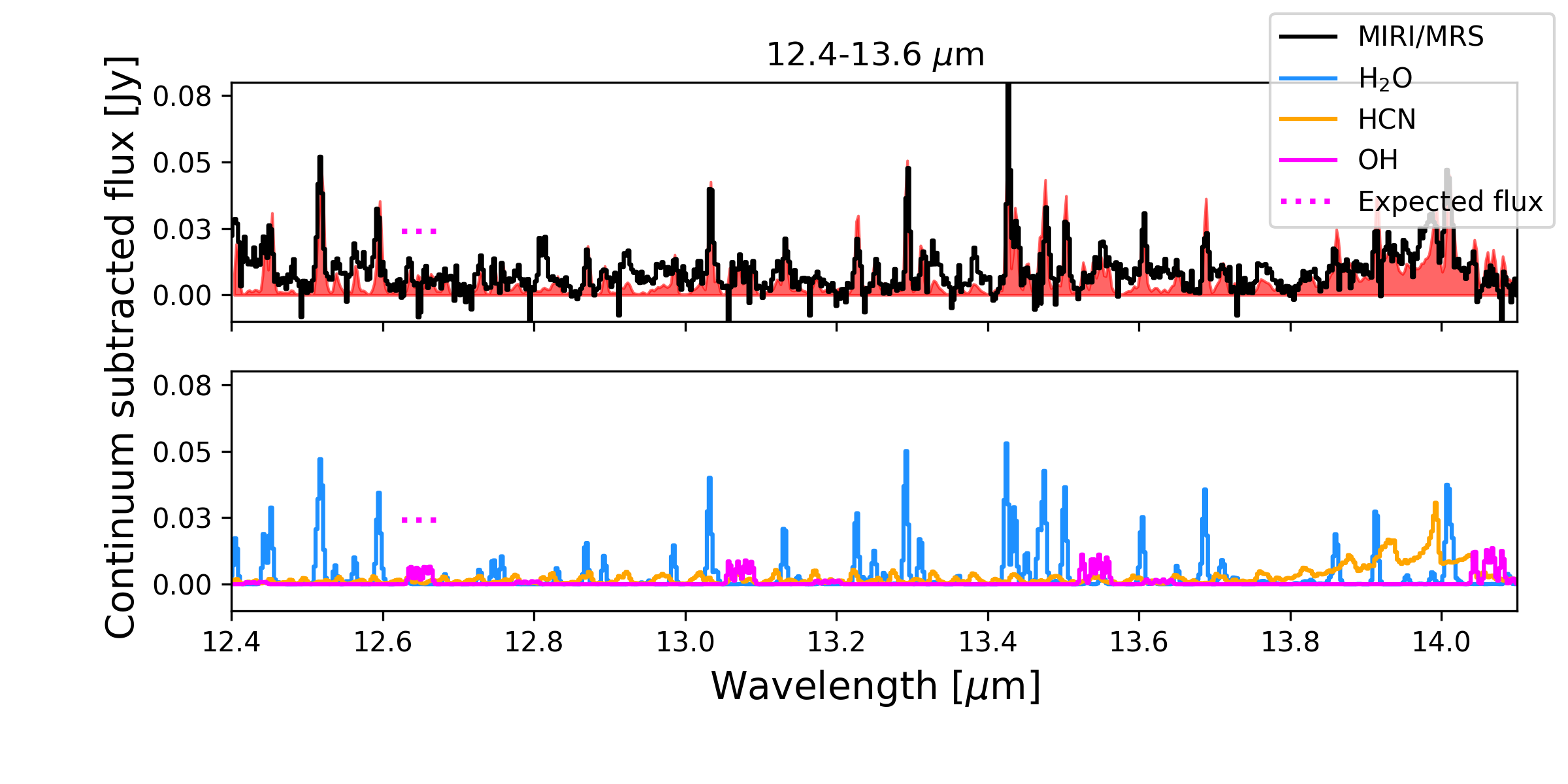}
    \caption{Close-up of the \ce{H2O} and \ce{OH} emission around $\sim$13~$\mu$m. The horizontal dotted line indicates the expected flux of the \ce{OH} quadruplet in case of prompt emission following \ce{H2O} photodissociation by Ly-$\alpha$ photons.}
    \label{fig:oh_h2o_zoom}
\end{figure*}

\ce{H2O} photodissociation by UV photons with a wavelength of $<$144~nm is known to produce \ce{OH} in high rotationally excited states \citep{ref:00HaHwYa,ref:00HaHe}. \citet{ref:21TaHeDi} show that newborn \ce{OH} formed by \ce{H2O} photodissociation produces a series of rotationally excited lines detectable longwards of 9~$\mu$m, a process called prompt emission. In the MIRI/MRS, this process can be traced by highly excited lines shortwards of 10~$\mu$m that can uniquely be excited by \ce{H2O} photodissociation. \citet{ref:21TaHeDi} show that the series of rotational lines of \ce{OH} exhibit relatively constant photon flux. Based on the strength of our detected \ce{OH} lines longwards of 14~$\mu$m, the lines in the 10~$\mu$m would be detected, but already around 12.65~$\mu$m the lines are much weaker. Figure \ref{fig:oh_h2o_zoom}, shows that only very weak, if any, \ce{OH} emission is found in this wavelength region. In case of prompt emission, it is expected that the number of photons is conserved down the rotational excitation ladder. The higher energy lines in the shorter wavelengths should therefore be brighter than the lower energy lines. Based on the flux of the quadruplet around $\sim$15.3~$\mu$m, it is expected that the flux of the quadruplet at $\sim$12.65~$\mu$m would be $\frac{15.3}{12.65}\approx 1.2$ times higher. This does not hold here, since these expected brighter lines are not observed. Additionally, prompt emission results in an asymmetry of the \ce{OH} quadruplets, which is not found in our MIRI/MRS data (\citealt[][Zannese et al., in prep.]{ref:14CaNa,ref:15ZhXiGu}). We therefore conclude that \ce{OH} lines are not primarily excited by \ce{H2O} photodissociation. The detected \ce{OH} lines are more likely excited by collision or by chemical pumping through the \ce{O + H2 -> OH + H} reaction. The non-detection of \ce{OH} prompt emission could indicate that small dust grains, which could be blocking the emission of the \ce{H2O}-flux from deeper layers, could also be strongly attenuating the UV radiation field.




\subsection{Reduced CO$_2$}
Similar to \ce{H2O}, the precursor for \ce{CO2} formation is \ce{OH}. However, this gas reaction is favoured over the formation of \ce{H2O} for $T \leq 250$~K \citep{ref:22BoBeCab}. In Table \ref{tab:chi_results} we find the excitation temperature for \ce{CO2} to be in this range, while the excitation temperatures for \ce{H2O} are higher in most regions. A lack of \ce{OH} in the deeper and colder layers of the disk due to UV shielding of \ce{H2O}, could prevent the formation of \ce{CO2} since it is expected to form in the deeper part of the disk.

\citet{ref:22BoBeCab} find that the self-shielding of \ce{H2O} inhibits the formation of \ce{CO2} due to a lack of \ce{OH}, while the abundance of \ce{CO2} can still be reduced due to dissociation deeper into the disk due to a lack of self-shielding. This is in contrast with other modelling efforts, where the \ce{CO2} $Q$-branch is generally overproduced compared to the spectrum of Sz~98 \citep[e.g.][]{ref:18WoMiTh,ref:21AnBlCl}. The temperature structure of the disk is critical: following \citet{ref:15GlNa}, when \citet{ref:22BoBeCab} add additional chemical heating in their work, the thermochemical equilibrium tips towards formation of \ce{H2O}, since this reaction is favoured for higher temperatures. These reasons combined (a change in temperature structure and self-shielding) result in a significantly reduced line flux for \ce{CO2}. As a result, the column density of the \ce{CO2} emission is expected to be $\sim$10$^{16}$~cm$^{-2}$ (which is slightly lower than expected from other models, \citealt{ref:21AnBlCl}), at lower temperatures. As presented in Table \ref{tab:chi_results}, we indeed find these lower temperatures, although the column density is higher. However, based on the confidence in the $\chi^2$ map due to the degeneracy between column density and emitting area presented in Fig. \ref{fig:chi2_region2}, the column density could very well be lower, and perhaps in the $\sim$10$^{16}$~cm$^{-2}$ range if the emission was more extended. It is therefore not certain that this discrepancy is truly present.

While the \ce{CO2} abundance in the inner regions of T Tauri protoplanetary disks can be greatly reduced due to the aforementioned mechanisms, \citet{ref:22BoBeCab} still find a relatively bright $Q$-branch compared to the \ce{H2O} emission when only including \ce{H2O} self-shielding in their models, which does not match what we find in the disk of Sz~98. In their base model, \ce{CO2} emits more strongly from cold gas farther out. When they assume that the emission for both species is contained within the warmer regions inside the \ce{H2O} snow line, a similar ratio of line-strengths is found as for Sz~98, indicating that the \ce{CO2}- and \ce{H2O}-emission is contained in a radially less extended region of the inner disk. \citet{ref:22BoBeCab} attribute this to physical processes being active, for example the 'cold-finger effect'. In their view, this would lock oxygen in the form of \ce{H2O}-ice at the midplane \ce{H2O} ice line, reducing the amount of oxygen available for \ce{CO2} formation and limiting its extent. However, as stated previously, far-IR lines would provide better constraints related to this. Based on Table \ref{tab:chi_results} for \ce{H2O} and the $Q$-branch-only fit in Fig. \ref{fig:chi2_region2} for region 2, both \ce{H2O} and \ce{CO2} seem to be contained with a similar radial extent. However, the $R$ value is related to an emitting area, not the location. The emission could therefore also originate from a ring further out in the disk, for example due to a small cavity. A theoretical exploration of the \ce{H2O} versus \ce{CO2} emission variations using thermochemical disk models will be presented in Vlasblom et al. (in prep.).


\subsection{Lack of other organics}
Due to the shielding of \ce{H2O} either due to itself, or small dust, less atomic oxygen and \ce{OH} are available for the formation of both \ce{CO2} and \ce{CO}. This means that the carbon is free to form hydrocarbons and other organic molecules \citep{ref:22DuBoBe}. However, aside from \ce{HCN} and \ce{CO2}, these species are not detected in the spectrum of Sz~98. GW~Lup, as presented by \citet{ref:23GrDiTa}, shows detectable \ce{C2H2}, which is lacking in the spectrum of Sz~98. After subtracting the best-fit slab models, we find an upper limit for \ce{C2H2} in the 13.65-13.72~$\mu$m region of $\sim$0.3$\times 10^{-14}$~erg~s$^{-1}$~cm$^{-2}$ normalised to a distance of 140~pc. As can be seen in \citet{ref:23TaBeDi}, this is among the lowest \ce{C2H2} fluxes in the sample of \citet{ref:20BaPaBo}. When detected, \ce{C2H2} has been fit with high temperatures in the past \citep{ref:11SaPoBl}, indicating it is located higher in the disk or at smaller radii. If the former, it is simply not abundant enough to be detected in Sz~98. On the other hand, modelling efforts expect \ce{C2H2} to lie deeper in the disk than \ce{HCN} and \ce{CO2} \citep{ref:18WoMiTh}. It is therefore possible that the organics are present, but blocked by the small dust deeper down in the inner disk. However, \ce{C2H2} was detected in the MIRI/MRS spectrum of EX~Lup \citep{ref:23KoAbDi}, despite the dust likely having a slightly smaller grain size based on the results in \citet{ref:06KeAuDu} and \citet{ref:23KoAbDi}, indicating more opaque dust.


On the other hand, the gas in the outer disk of Sz~98 has been inferred to have a C/O$>$1 \citep{ref:19MiFaVa}, whereas the inner disk may be poor in carbon to begin with. The infall of icy pebbles is expected to mainly replenish the inner disk's oxygen abundance rather than its volatile carbon abundance \citep[e.g.][]{ref:21ObBe}, resulting in a lower volatile elemental C/O ratio than expected from observations of the outer disk. The diversity in the \ce{C2H2} and \ce{HCN} fluxes in \textit{Spitzer} samples could be caused by differing C/O ratios between disks \citep{ref:11CaNa,ref:15WaNoDi}. The larger the deviation from solar C/O ratios ($\sim$0.5), the larger the range in flux ratios. The samples examined in \citet{ref:11CaNa} (line fluxes: \ce{H2O}, \ce{CO2}, \ce{HCN}) and \citet{ref:11SaPoBl} (line fluxes: \ce{CO2}, \ce{HCN}, \ce{OH}, \ce{C2H2}) have line flux ratios spanning from 0.1 to 10. However, the lack of (or at the very least extremely weak) \ce{C2H2} and \ce{CH4} in Sz~98 indicates that a larger deviation from solar C/O is to be expected. According to Fig. 11 of \citet{ref:21AnBlCl}, this could correspond more to a C/O of approximately 0.14 or lower, depending on the other physical properties of the inner disk. They find the \ce{HCN/H2O} flux ratio to be a good tracer for the C/O ratio in the inner disk. We use the same region of the spectrum to calculate this flux ratio, focusing on the $\sim$13.9~$\mu$m and $\sim$17.23~$\mu$m lines of \ce{HCN} and \ce{H2O}, respectively. After subtracting the slab model fits presented in Fig. \ref{fig:spec_ranges} to get a `clean' \ce{HCN} feature, the \ce{HCN/H2O} flux ratio is $\sim$1.7. Assuming their fiducial model and that this is only caused by the C/O ratio, this would indeed indicate that C/O$<$0.14 \citep{ref:21AnBlCl}. This is in contrast with the observations of \citet{ref:13NaCaPo}, where the larger disk mass should result in a larger C/O ratio. They link this to the idea that disks of higher mass could more readily form larger planetesimals, depleting the inner disk of gaseous \ce{H2O}. It could therefore be the case that this is not true for Sz~98.

However, the line fluxes may change when altering the fiducial model presented in \citet{ref:21AnBlCl}. In order to investigate this, \citet{ref:21AnBlCl} varied a selection of parameters in their models, and we discuss some of their conclusions in relation to the observations here. For example, a larger inner gas radius of 0.5~au, similar to what has been suggested for the dust cavity of Sz~98 \citep{ref:19TeDiCa}, instead of 0.2~au reduces the line fluxes of most species (and would reduce the line flux of \ce{H2O} at shorter wavelengths, see e.g. \citealt{ref:17BaPoSa,ref:23BaPoPe}), but the influence on more extended species is limited (e.g. \ce{C2H2} and \ce{OH}). A larger inner gas radius is therefore unlikely in Sz~98, since the relative line fluxes lean towards stronger \ce{H2O} and \ce{HCN} instead, both of which are found to be less extended \citep{ref:21AnBlCl}. On the other hand, when assuming the gas temperature is similar to the dust temperature, \citet{ref:21AnBlCl} find a lower line flux for \ce{C2H2} and \ce{OH}, while the other molecules are unchanged. This could also cause the larger difference in line flux ratios, allowing the C/O ratio to be slightly higher than 0.14. However, the \ce{HCN}/\ce{H2O} is less sensitive to this change. On the other hand, based on ProDiMo models \citet{ref:23AnKaWa} find the \ce{HCN}/\ce{H2O} to be more sensitive to the dust opacity, where the ratio decreases for less opaque dust. While this may be part of the cause why the flux ratio is relatively low, it is still possible that the C/O is much lower than the solar value, limiting the abundances of hydrocarbons.

In App. \ref{app:alma} the \ce{CO} snow line is estimated to lie at $\sim$20~au, which lies in the first ring located at $\sim$27~au separated from the inner disk by the gap at $\sim$16~au. This gap could block \ce{CO} from reaching the inner disk. It is therefore possible that \ce{H2O}-ice migrated inwards first, reaching the inner disk, sublimating, and replenishing the oxygen reservoir prior to formation of the gaps; while \ce{CO}-ice got trapped in the ring before it could migrate further. This would prevent more \ce{CO} from reaching the inner disk, where a larger amount of oxygen would already be present from sublimated \ce{H2O}-ice, therefore reducing the C/O ratio. The timescale over which the different ices are delivered to the inner disk compared to the timescale over which the gaps formed proves crucial to understanding the composition of the inner disk. Combining studies discussing relative abundances of gas and ice species over time such as \citet{ref:18EiWaDi,ref:22EiHe} with gap formation could shed more light on the effects on the species available to accreting planets in the inner disk.

\section{Conclusions}
\label{sec:conclusion}
We presented the \textit{JWST} MIRI/MRS spectrum of the inner disk of Sz~98, a disk previously only observed with the low-resolution mode of \textit{Spitzer} IRS. The improved resolution and sensitivity of the MRS reveal a rich spectrum full of both ro-vibrational and pure rotational \ce{H2O} lines superposed on the continuum. Aside from \ce{H2O}, we detect \ce{CO}, \ce{CO2}, \ce{HCN}, and \ce{OH}. Despite the disk's large size, the thick \ce{H2O} column indicates that grains have likely drifted in towards the star, allowing \ce{H2O} ice to sublimate, forming an optically thick, potentially self-shielding layer of gaseous \ce{H2O}. Additionally, the spectrum likely probes different \ce{H2O} reservoirs from the inner parts outwards, when analysing data from shorter to longer wavelengths. This property must be considered in future work when fitting the entire MIRI/MRS spectral range. The line fluxes at longer wavelengths are still quite high, indicating that most of the \ce{H2O} is present at the surface, and could even be past the mid-plane snow line.

We find several signs pointing towards limited photodissociation of \ce{H2O}, due to self-shielding and/or dust extinction. First, its column density is larger than the few times 10$^{17}$~cm$^{-2}$ required for self-shielding. Second, the \ce{H2O} line fluxes are relatively low, potentially due the presence of small dust blocking the deeper parts of the \ce{H2O} column and causing the extinction of the UV flux. Finally, the \ce{CO2} emission is relatively weak in this disk, due to limited availability of \ce{OH} for its formation, locking most of the oxygen up in \ce{H2O}.

The lack of other organic molecules, most notably \ce{C2H2}, in the spectrum of the inner disk of Sz~98 is indicative of a low volatile elemental C/O ratio, potentially 0.14 or lower. The use of line fluxes to determine the C/O ratio has drawbacks, since they are dependent on other disk properties as well. However, a sub-solar C/O ratio seems likely for the Sz~98 inner disk, while the outer disk exhibits a high C/O and low C/H. Sz~98 is not unique in this regard; many of the disks observed with \textit{Spitzer} show similar \ce{H2O} and \ce{OH/H2O} line fluxes, although for the inner disk of Sz~98 the line fluxes are on the lower end compared to the sample in \citet{ref:17BaPoSa}. If organics are also lacking in the other disks in the same range, this could mean their C/O ratios are similarly low. The timescale over which ices migrate versus the timescale over which substructures form likely influences the C/O ratio of the inner disk.

\begin{acknowledgements}
D.G. thanks John H. Black for his insightful comments.

This work is based on observations made with the NASA/ESA/CSA James Webb Space Telescope. The data were obtained from the Mikulski Archive for Space Telescopes at the Space Telescope Science Institute, which is operated by the Association of Universities for Research in Astronomy, Inc., under NASA contract NAS 5-03127 for JWST. These observations are associated with program \#1282 and \#1050.

The following National and International Funding Agencies funded and supported the MIRI development: NASA; ESA; Belgian Science Policy Office (BELSPO); Centre Nationale d’Etudes Spatiales (CNES); Danish National Space Centre; Deutsches Zentrum fur Luftund Raumfahrt (DLR); Enterprise Ireland; Ministerio De Econom\'ia y Competividad; Netherlands Research School for Astronomy (NOVA); Netherlands Organisation for Scientific Research (NWO); Science and Technology Facilities Council; Swiss Space Office; Swedish National Space Agency; and UK Space Agency.

This paper makes use of the following ALMA data: ADS/JAO.ALMA\#2018.0.01458.S. ALMA is a partnership of ESO (representing its member states), NSF (USA) and NINS (Japan), together with NRC (Canada), MOST and ASIAA (Taiwan), and KASI (Republic of Korea), in cooperation with the Republic of Chile. The Joint ALMA Observatory is operated by ESO, AUI/NRAO and NAOJ.

This work has made use of data from the European Space Agency (ESA) mission
{\it Gaia} (\url{https://www.cosmos.esa.int/gaia}), processed by the {\it Gaia}
Data Processing and Analysis Consortium (DPAC,
\url{https://www.cosmos.esa.int/web/gaia/dpac/consortium}). Funding for the DPAC
has been provided by national institutions, in particular the institutions
participating in the {\it Gaia} Multilateral Agreement.

A.C.G. has been supported by PRIN-INAF MAIN-STREAM 2017 “Protoplanetary disks seen through the eyes of new- genera-tion instruments” and from PRIN-INAF 2019 “Spectroscopically tracing the disk dispersal evolution (STRADE)”.

G.B. thanks the Deutsche Forschungsgemeinschaft (DFG) - grant 138 325594231, FOR 2634/2.

E.v.D. acknowledges support from the ERC grant 101019751 MOLDISK and the Danish National Research Foundation through the Center of Excellence ``InterCat'' (DNRF150). 

D.G. would like to thank the Research Foundation Flanders for co-financing the present research (grant number V435622N).

T.H. and K.S. acknowledge support from the European Research Council under the Horizon 2020 Framework Program via the ERC Advanced Grant Origins 83 24 28. 

I.K., A.M.A., and E.v.D. acknowledge support from grant TOP-1 614.001.751 from the Dutch Research Council (NWO).

I.K. and J.K. acknowledge funding from H2020-MSCA-ITN-2019, grant no. 860470 (CHAMELEON).

B.T. is a Laureate of the Paris Region fellowship program, which is supported by the Ile-de-France Region and has received funding under the Horizon 2020 innovation framework program and Marie Sklodowska-Curie grant agreement No. 945298.

O.A. and V.C. acknowledge funding from the Belgian F.R.S.-FNRS.

I.A., D.G. and B.V. thank the Belgian Federal Science Policy Office (BELSPO) for the provision of financial support in the framework of the PRODEX Programme of the European Space Agency (ESA).

L.C. acknowledges support by grant PIB2021-127718NB-I00,  from the Spanish Ministry of Science and Innovation/State Agency of Research MCIN/AEI/10.13039/501100011033.

T.P.R acknowledges support from ERC grant 743029 EASY.

D.R.L. acknowledges support from Science Foundation Ireland (grant number 21/PATH-S/9339).

D.B. has been funded by Spanish MCIN/AEI/10.13039/501100011033 grants PID2019-107061GB-C61 and No. MDM-2017-0737.

M.T. acknowledges support from the ERC grant 101019751 MOLDISK

\end{acknowledgements}

\bibliography{bib} 
\bibliographystyle{aa} 

\begin{appendix}

\section{$\chi^2$ maps of fits}
\label{app:chi2}
\begin{figure*}
\centering
\begin{subfigure}
  \centering
  \includegraphics[width=.49\linewidth]{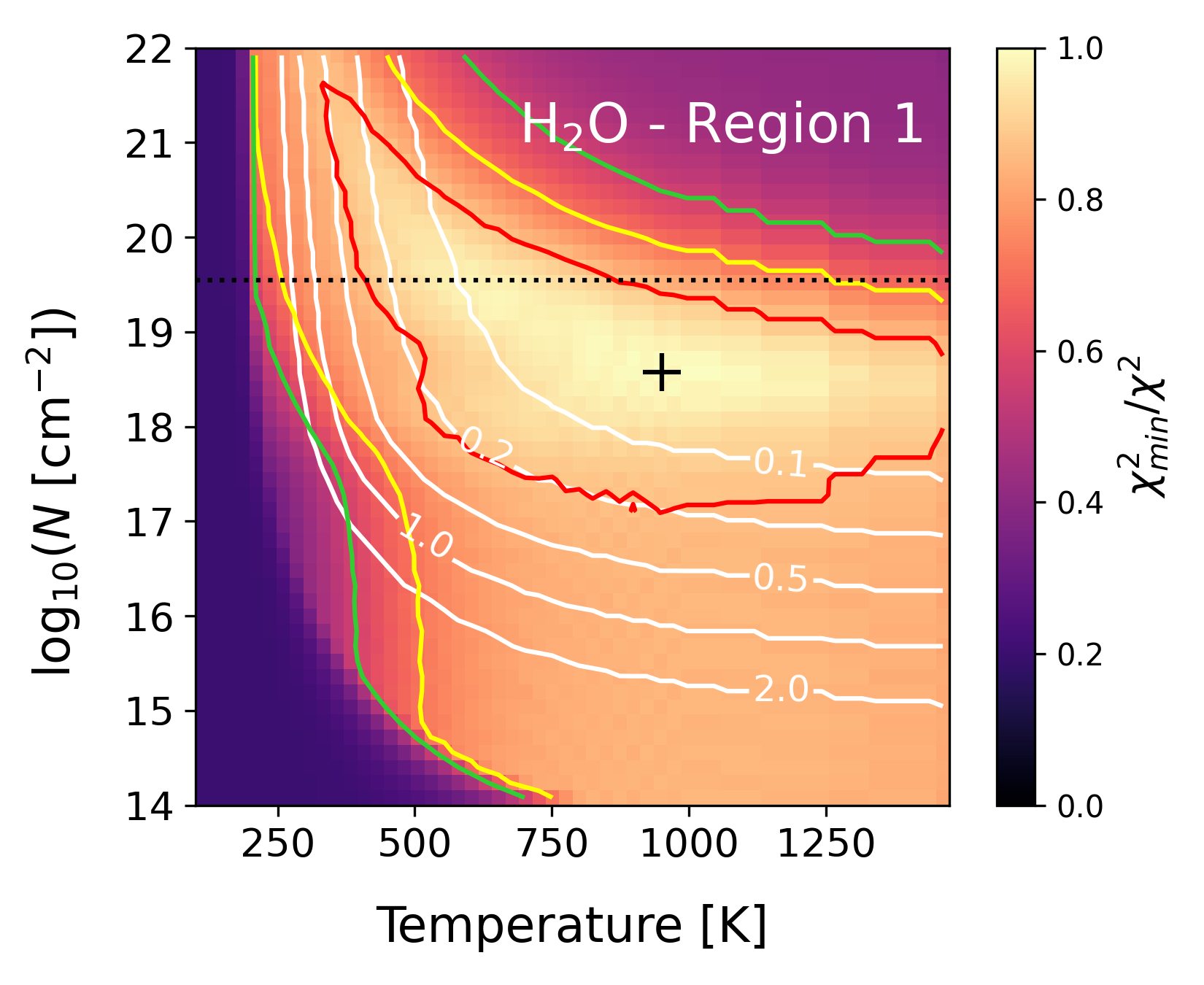}
\end{subfigure}%
\begin{subfigure}
  \centering
  \includegraphics[width=.49\linewidth]{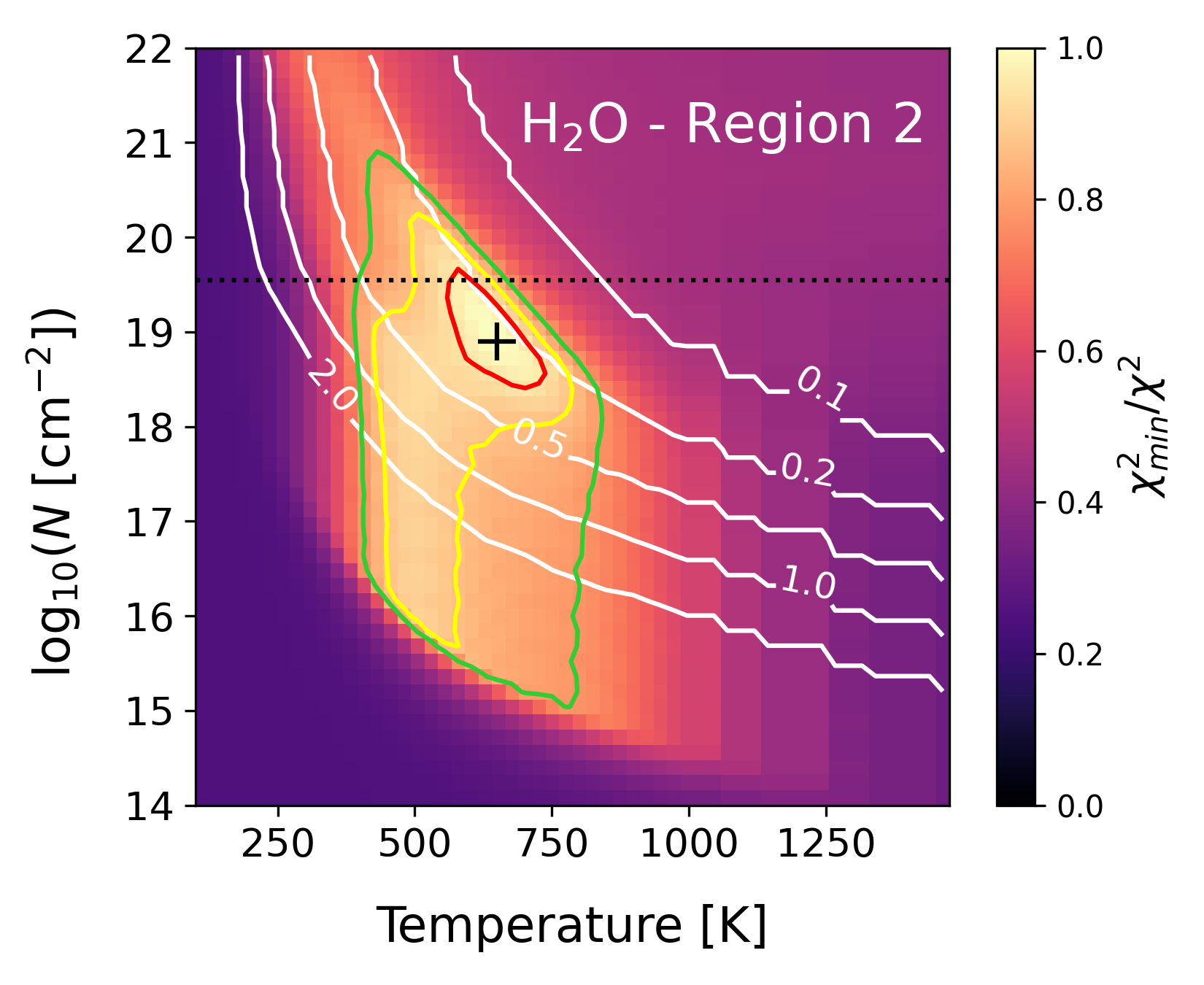}
\end{subfigure}

\begin{subfigure}
  \centering
  \includegraphics[width=.49\linewidth]{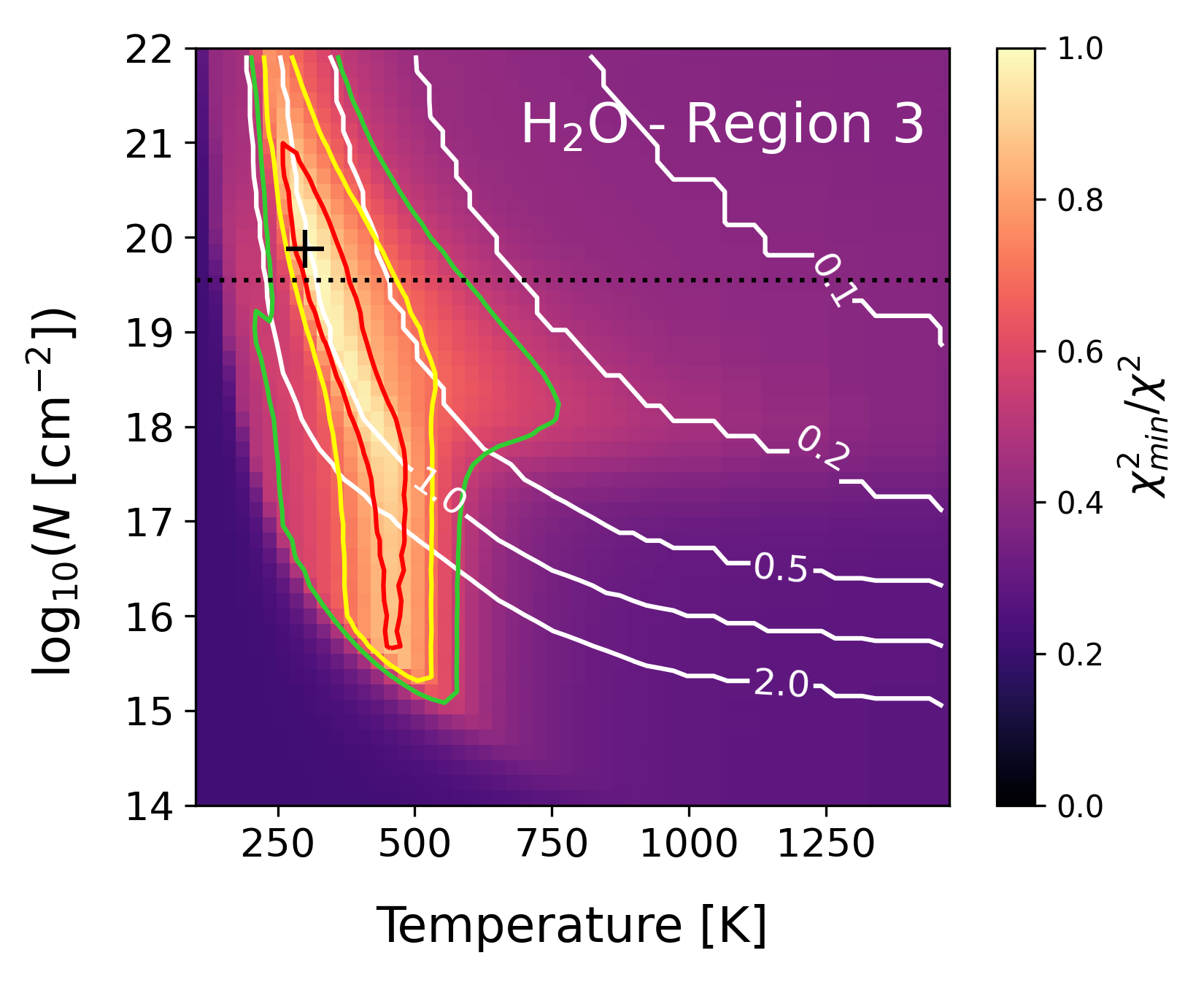}
\end{subfigure}%
\begin{subfigure}
  \centering
  \includegraphics[width=.49\linewidth]{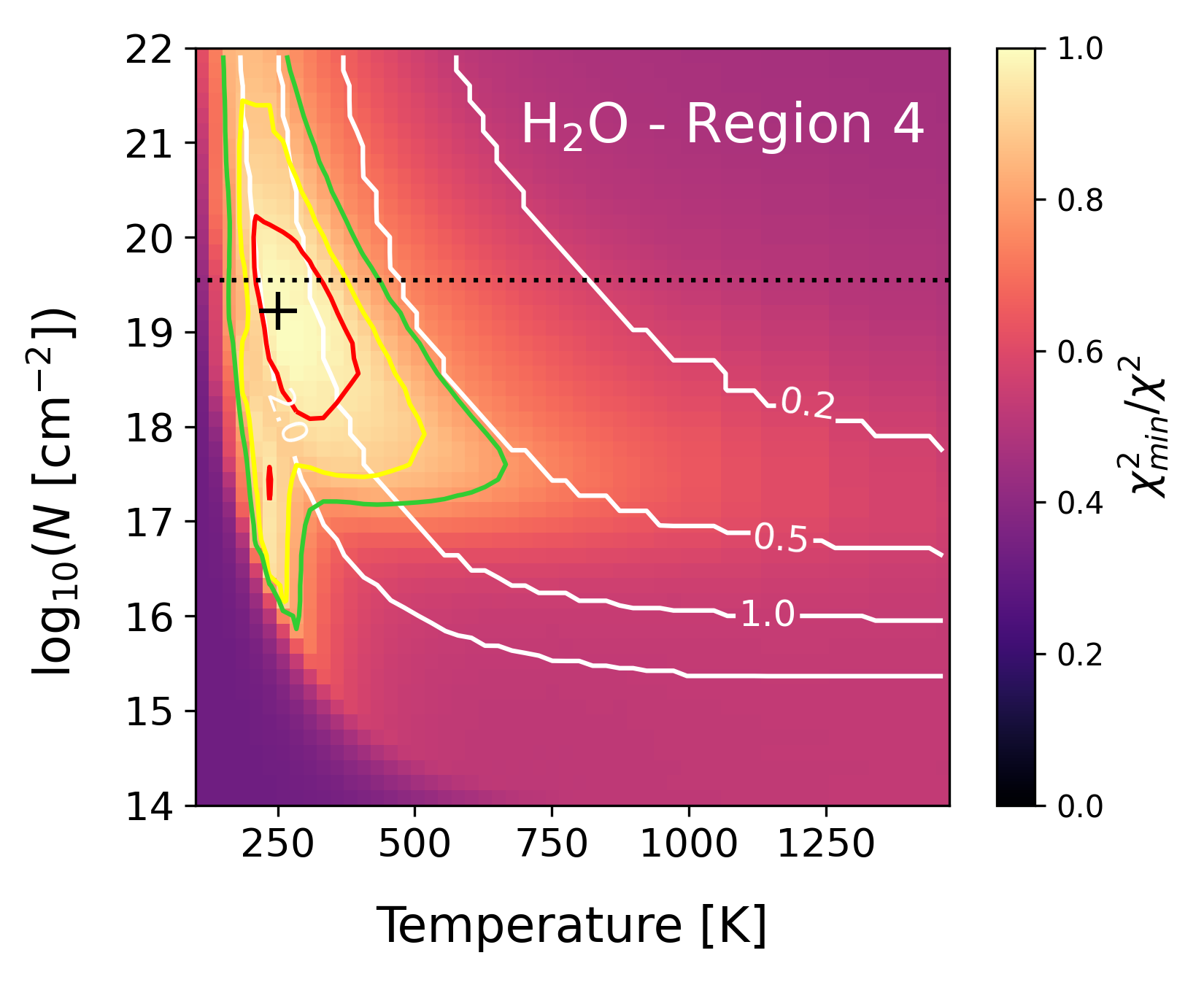}
\end{subfigure}
\caption{$\chi^2$ plots of \ce{H2O} over the different regions of the disk (top left: region 1, top right: region 2, bottom left: region 3, bottom right: region 4). The red, yellow, and green lines indicate the $1\sigma$, $2\sigma$, and $3\sigma$ confidence contours, respectively. The white contours show the emitting radii in astronomical units (0.1 to 2.0 au). The black cross corresponds to the best fit. The black dotted line corresponds to the upper limit of the column density based on the H$_\text{2}^{\text{18}}$O and H$_\text{2}^{\text{16}}$O flux ratio as described in Sect. \ref{subsec:water}.}
\label{fig:chi2_h2o}
\end{figure*}

\begin{figure*}
\centering
\begin{subfigure}
  \centering
  \includegraphics[width=.49\linewidth]{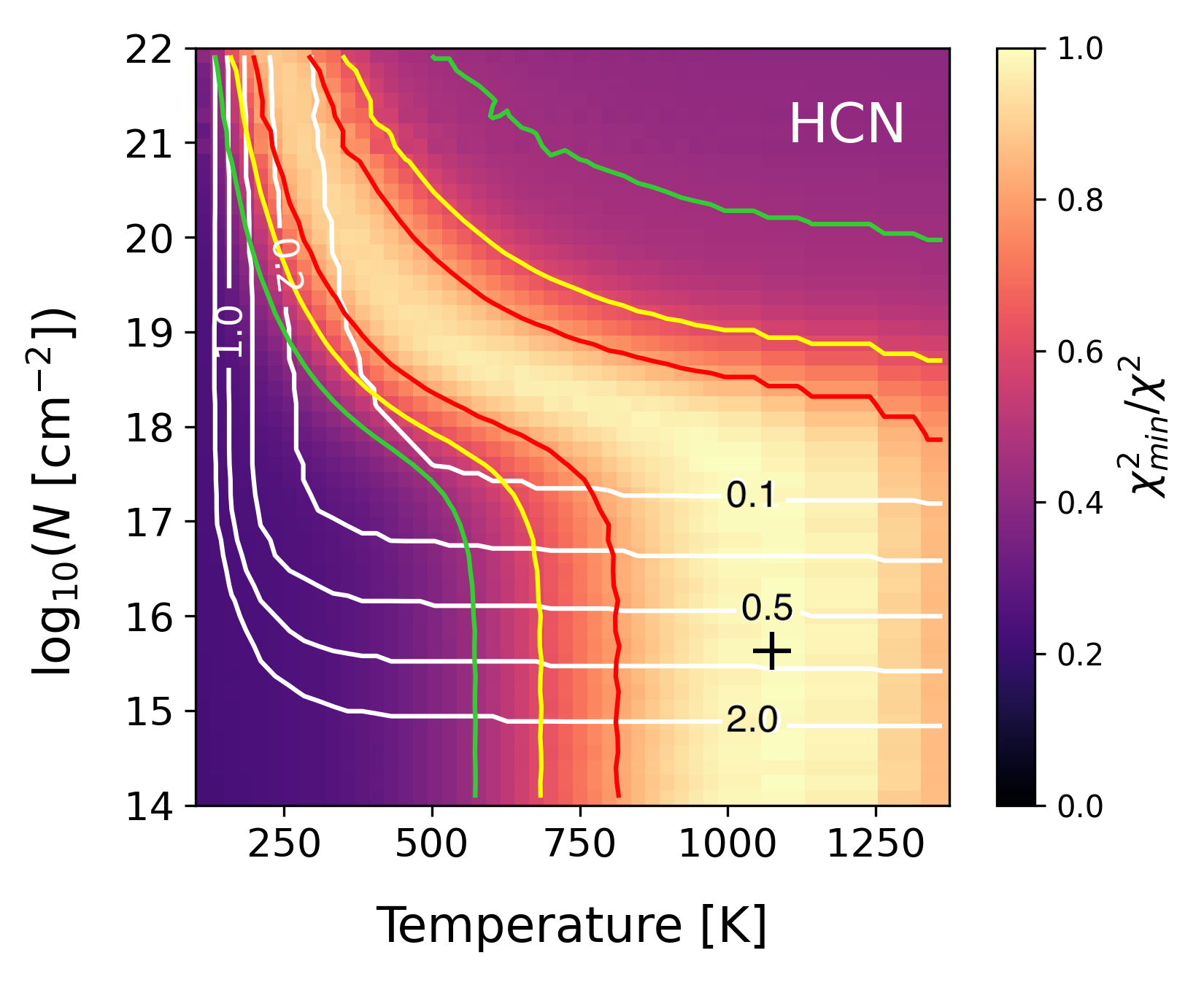}
\end{subfigure}%
\begin{subfigure}
  \centering
  \includegraphics[width=.49\linewidth]{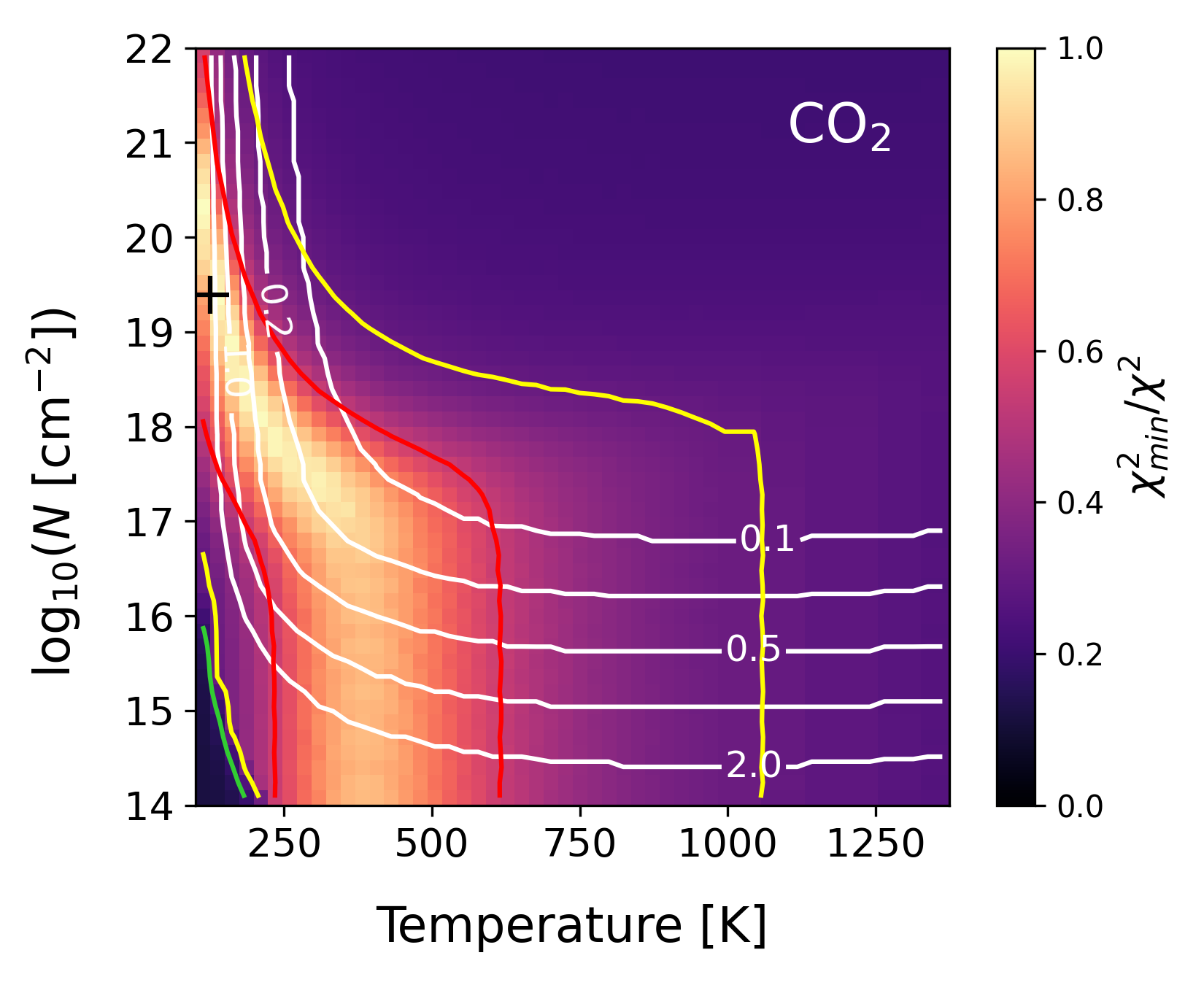}
\end{subfigure}
\begin{subfigure}
  \centering
  \includegraphics[width=.49\linewidth]{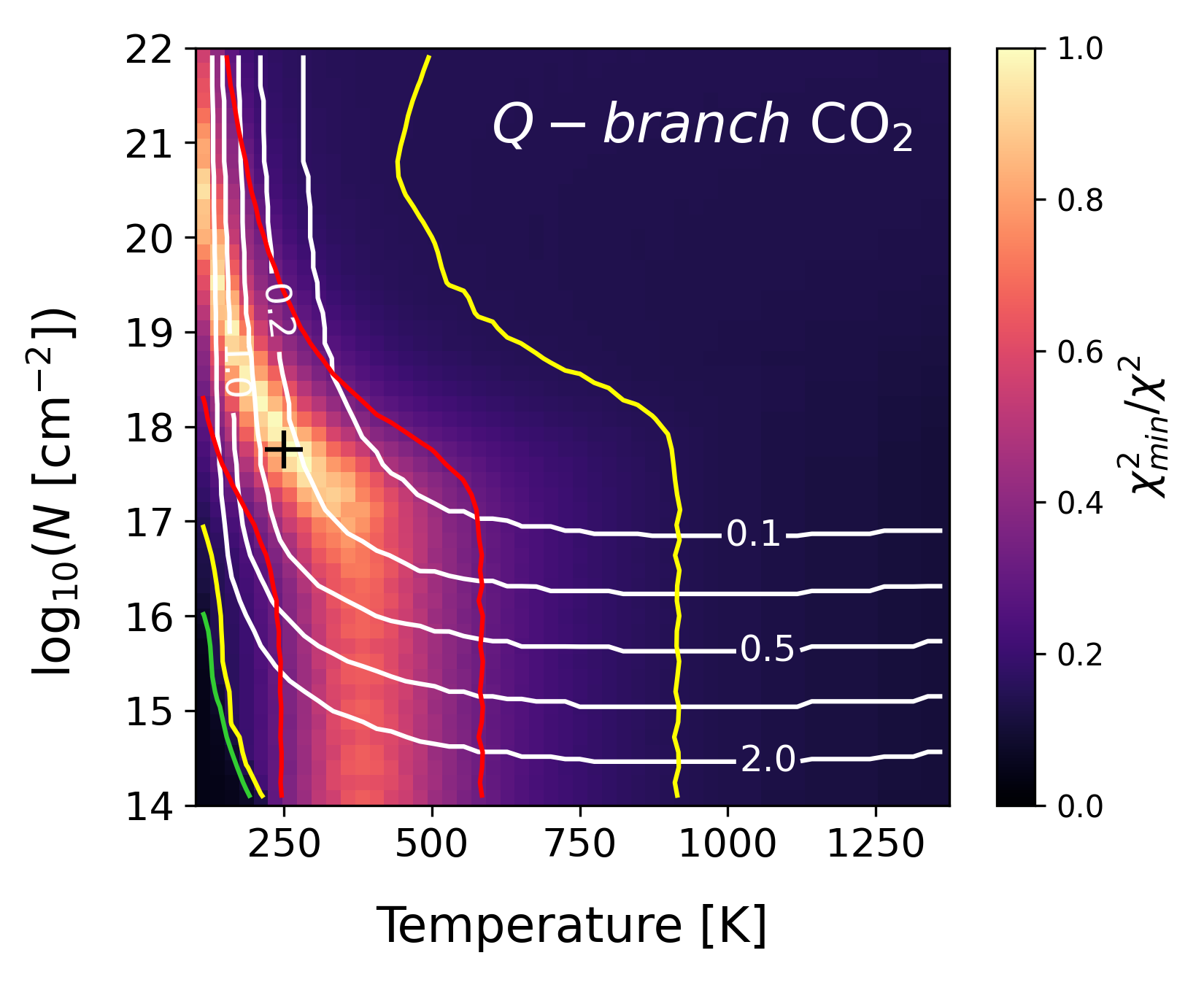}
\end{subfigure}

\caption{$\chi^2$ plots of \ce{HCN} and \ce{CO2} detected in region 2. The red, yellow, and green lines indicate the $1\sigma$, $2\sigma$, and $3\sigma$ confidence contours, respectively. The white contours show the emitting radii in astronomical units (0.1 to 2.0 au). The black cross corresponds to the best fit. The bottom plot shows the map for \ce{CO2} when fitting the $Q$-branch only.}
\label{fig:chi2_region2}
\end{figure*}

The $\chi^2$ maps per molecule and per region are presented here. \ce{H2O} across all regions is presented in Fig. \ref{fig:chi2_h2o}. The confidence of \ce{HCN} and \ce{CO2} detected in region 2 can be found in Fig. \ref{fig:chi2_region2}.

\section{Masked features}
\label{app:blank}
The wavelengths at which spurious data reduction artefacts have been masked are:
[5.0091,5.01071]; [5.018,5.019]; [5.112,5.15]; [5.2157,5.2184]; [5.2267,5.2290]; [5.2441,5.2471]; [5.2947,5.2974]; [5.3742,5.3777]; [5.3836,5.3877]; [5.4181,5.4210]; [5.5644,5.5674]; [5.5925,5.5966]; [5.8252,5.8267]; [5.8669,5.8689]; [5.9,5.916]; [5.9282,5.9314]; [5.9691,5.9728]; [6.0357,6.0394]; [6.0430,6.0462]; [6.1012,6.1044]; [6.1311,6.1421]; [6.26,6.31]; [6.3740,6.3757]; [6.3783,6.3810]; [18.8055,18.8145]; [19.004,19.012]; [21.974,21.985]; [25.69824,25.71313].

Due to the calibration files being based on an A~star, some of its features are propagated into the target here. The majority of these features are present in the short wavelengths, resulting in most of the blanked lines being present here.

\section{CO$_2$ column density}
\label{app:co2}
\begin{figure*}
    \centering
    \includegraphics[width=\textwidth]{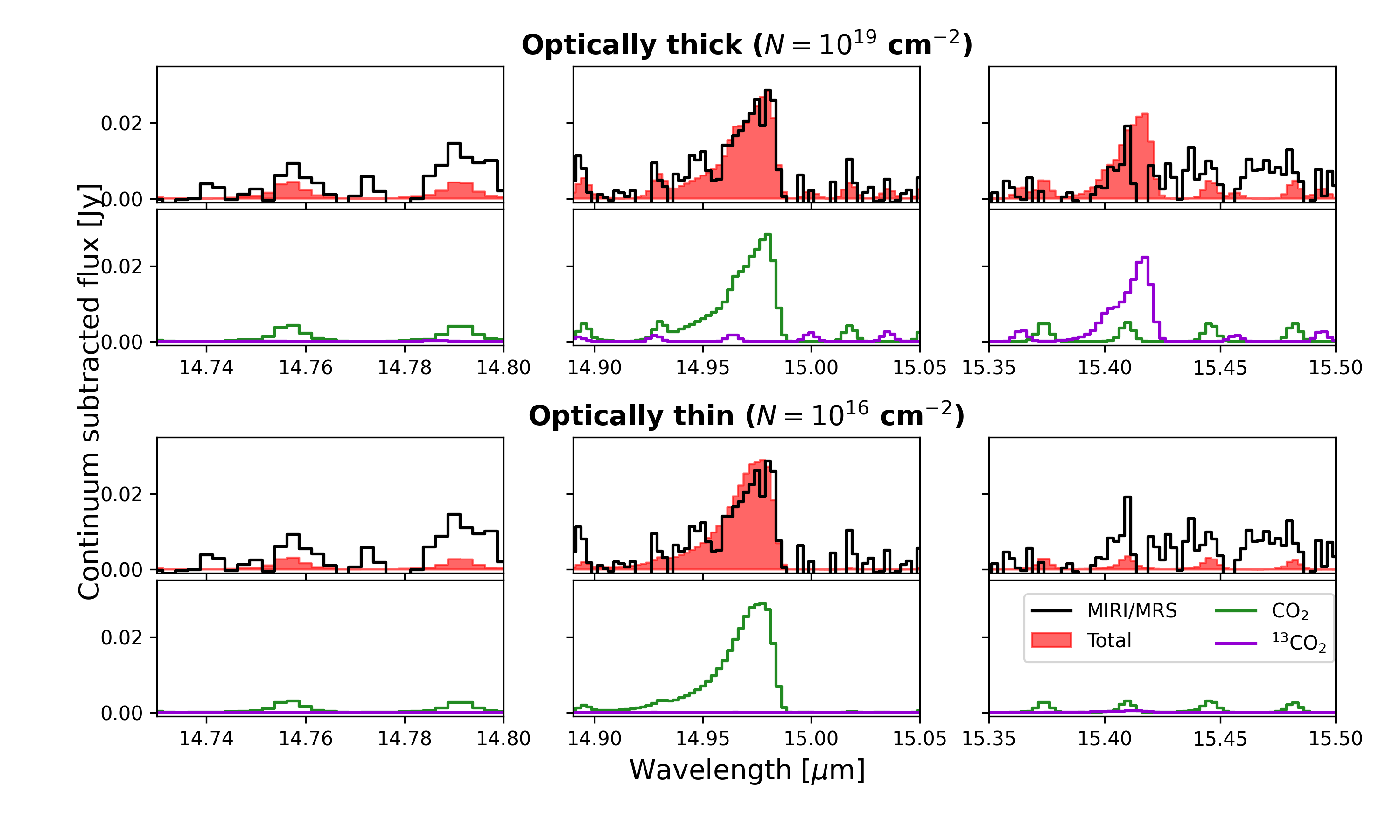}
    \caption{Comparison between the optically thick and thin cases for \ce{CO2} and $^{\text{13}}$CO$_\text{2}$ assuming a ratio of 68, overlain on the data with \ce{H2O}, \ce{HCN}, and \ce{OH} subtracted. Due to the faintness of the features and blending with \ce{H2O} lines it becomes difficult to constrain the best model.}
    \label{fig:co2_slab}
\end{figure*}

The column density of \ce{CO2} is difficult to constrain due to the faintness of the features and the overlap with \ce{H2O}. In Fig. \ref{fig:co2_slab} it can be seen that distinguishing between the optically thin and thick cases is non-trivial.

\section{Sz~98 from ALMA data}
\label{app:alma}
The continuum image shown in Fig. \ref{fig:continuum} has been created using the continuum spectral window (centred at 232.984~GHz, 1286.75~$\mu$m) contained in the ALMA archival data set 2018.1.01458.S (PI: Hsi-Wei Jen). The data set was calibrated using the provided pipeline scripts and the specified \textit{Common Astronomy Software Applications} (\textit{CASA}) \citep{ref:07McWaSc}. The continuum image was created using \textit{CASA} version 6.5.2.26 and the \textsc{TCLEAN}-task, where we have used the Briggs weighting scheme and a robust parameter of +1.0. We cleaned down to a threshold of $\sim1\times$ the RMS in the initial, dirty image. The inclination of the disk is 47.1$\degree$ with a position angle of 111.6$\degree$ \citep{ref:17TaTeNa}.

The cleaned image has a resolving beam of 0.079"$\times$0.063" (-27.135$\degree$). The emission peaks at 2.167~mJy~beam$^{-1}$ and the image has a root mean square of 0.071~mJy~beam$^{-1}$. We have estimated the locations of the rings and gaps (displayed in Fig.~\ref{fig:continuum}) using a deprojected, azimuthally averaged radial profile. The radial profile was created using bin sizes of half the width of the beam's minor axis. As displayed in Fig. \ref{fig:continuum}, we infer the millimetre emission to consist of an inner compact core, surrounded by two rings at approximately 27 and 103 au, respectively. The gaps are expected to be located at approximately 16 and 88 au.

The location of the \ce{H2O}, \ce{CO2}, and \ce{CO} snow lines are estimated from a radiative transfer model that fits the spectral energy distribution, mm dust radial profiles and the \textit{JWST} continuum. Assuming that these molecules freeze-out at temperatures of, respectively, 150~K, 72~K, and 20~K \citep[e.g.][]{ref:04CoAnCh}, the model suggests that the H$_2$O snowline is located between approximately 1.05 and 1.11~au, the CO$_2$ snowline between approximately 2.17 and 2.29~au, and the CO snowline between approximately 20.64 and 21.40~au.

\end{appendix}

\end{document}